\documentclass[graybox]{svmult}

\usepackage{type1cm}        
\usepackage{makeidx}         
\usepackage{graphicx}        
\usepackage{multicol}        
\usepackage[bottom]{footmisc}
\usepackage{cite}

\makeindex             

\newcommand{\mycomment}[1]{{}}

\usepackage{amsmath,amssymb}
\usepackage{xspace}
\usepackage{bm}
\usepackage{bbm}

\newcommand{\set}[1]{{\mathcal{#1}}\xspace} 
\newcommand{\mat}[1]{{\mathbf{#1}}\xspace} 
\renewcommand{\vec}[1]{{\mathbf{#1}}\xspace} 

\newcommand{\parens}[1]{{\left(#1\right)}\xspace}
\newcommand{\brackets}[1]{{\left[#1\right]}\xspace}
\newcommand{\braces}[1]{{\left\{#1\right\}}\xspace}
\newcommand{\bars}[1]{{\left\vert#1\right\vert}\xspace}
\newcommand{\doublebars}[1]{{\left\Vert#1\right\Vert}\xspace}

\newcommand{\complex}{\ensuremath{\mathbb{C}}\xspace}

\newcommand{\floor}[1]{\ensuremath{\left\lfloor{#1}\right\rfloor}\xspace}

\newcommand{\card}[1]{\bars{#1}}

\newcommand{\setcomplex}{\ensuremath{\complex}}

\newcommand{\setmatrix}[3]{\ensuremath{#1^{#2 \times #3}}\xspace}

\newcommand{\setmatrixcomplex}[2]{\setmatrix{\setcomplex}{#1}{#2}}

\newcommand{\inv}{\ensuremath{^{-1}}\xspace}

\newcommand{\ctrans}{\ensuremath{^{{*}}}\xspace}

\newcommand{\entry}[2]{\ensuremath{\brackets{#1}_{#2}}\xspace}

\newcommand{\logtwo}[1]{\ensuremath{\mathrm{log}_{2}\parens{#1}}}

\newcommand{\diag}[1]{\ensuremath{\mathrm{diag}\parens{#1}}}

\newcommand{\angleop}[1]{\ensuremath{\mathrm{angle}\parens{#1}}}

\newcommand{\pnorm}[2]{\ensuremath{\doublebars{#2}_{#1}}\xspace}

\newcommand{\normtwo}[1]{\pnorm{2}{#1}}

\newcommand{\normfro}[1]{\pnorm{\mathrm{F}}{#1}}

\DeclareMathOperator*{\expect}{\mathbb{E}\xspace}

\DeclareMathOperator*{\argmin}{argmin}
\DeclareMathOperator*{\argmax}{argmax}

\newcommand{\maxop}[1]{\ensuremath{\mathrm{max}\parens{#1}}\xspace}

\newcommand{\st}{\ensuremath{\mathrm{s.t.~}}\xspace}
\newcommand{\opt}{\ensuremath{^{\star}}\xspace}

\newcommand{\powernoise}{\ensuremath{P_{\mathrm{noise}}}\xspace}
\newcommand{\powerdes}{\ensuremath{P_{\mathrm{des}}}\xspace}
\newcommand{\powerint}{\ensuremath{P_{\mathrm{int}}}\xspace}
\newcommand{\powertx}{\ensuremath{P_{\mathrm{tx}}}\xspace}

\newcommand{\powersi}{\ensuremath{P_{\mathrm{SI}}}\xspace}

\newcommand{\powertxue}{\ensuremath{\powertx^{\mathrm{UE}}}\xspace}
\newcommand{\powertxbs}{\ensuremath{\powertx^{\mathrm{BS}}}\xspace}

\newcommand{\snr}{\ensuremath{\mathsf{SNR}}\xspace}

\newcommand{\sinr}{\ensuremath{\mathsf{SINR}}\xspace}

\newcommand{\inr}{\ensuremath{\mathsf{INR}}\xspace}

\newcommand{\Atx}{\ensuremath{\mA_{\mathrm{tx}}}\xspace}
\newcommand{\Arx}{\ensuremath{\mA_{\mathrm{rx}}}\xspace}

\newcommand{\Nt}{\ensuremath{N_\mathrm{t}}\xspace} 
\newcommand{\Nr}{\ensuremath{N_\mathrm{r}}\xspace} 

\newcommand{\precb}{\ensuremath{\mathcal{F}}\xspace}

\newcommand{\comcb}{\ensuremath{\mathcal{W}}\xspace}

\newcommand{\nbr}{\ensuremath{\parens{\Delta\theta,\Delta\phi}}\xspace}
\newcommand{\nbrv}{\ensuremath{\parens{\Deltavartheta,\Deltavarphi}}\xspace}
\newcommand{\nbrd}{\ensuremath{\parens{\deltatheta,\deltaphi}}\xspace}
\newcommand{\nbrnbrd}{\ensuremath{\parens{\Deltatheta,\Deltaphi,\deltatheta,\deltaphi}}\xspace}
\newcommand{\nbrnbrdth}{\ensuremath{\parens{\Deltatheta,\deltatheta}}\xspace}
\newcommand{\nbrnbrdph}{\ensuremath{\parens{\Deltaphi,\deltaphi}}\xspace}
\newcommand{\nbrvnbrd}{\ensuremath{\parens{\Deltavartheta,\Deltavarphi,\deltatheta,\deltaphi}}\xspace}

\newcommand{\nbroneone}{\ensuremath{\parens{1^\circ,1^\circ}}\xspace}

\newcommand{\nbrtwotwo}{\ensuremath{\parens{2^\circ,2^\circ}}\xspace}

\newcommand{\thph}{\ensuremath{\parens{\theta,\phi}}\xspace}
\newcommand{\thphtx}{\ensuremath{\parens{\thetatx,\phitx}}\xspace}
\newcommand{\thphrx}{\ensuremath{\parens{\thetarx,\phirx}}\xspace}
\newcommand{\thphtxrx}{\ensuremath{\parens{\thetatx,\phitx,\thetarx,\phirx}}\xspace}

\newcommand{\thphtxiopt}{\ensuremath{\parens{\thetatx\idx{i\opt},\phitx\idx{i\opt}}}\xspace}
\newcommand{\thphrxjopt}{\ensuremath{\parens{\thetarx\idx{j\opt},\phirx\idx{j\opt}}}\xspace}

\newcommand{\thphtxopt}{\ensuremath{\parens{\thetatx\opt,\phitx\opt}}\xspace}
\newcommand{\thphrxopt}{\ensuremath{\parens{\thetarx\opt,\phirx\opt}}\xspace}

\newcommand{\idx}[1]{\ensuremath{^{\parens{#1}}}\xspace}

\newcommand{\txdirsetmeas}{\ensuremath{\set{T}}\xspace}
\newcommand{\rxdirsetmeas}{\ensuremath{\set{R}}\xspace}

\newcommand{\txdirsetcb}{\ensuremath{\set{A}_{\mathrm{tx}}}\xspace}
\newcommand{\rxdirsetcb}{\ensuremath{\set{A}_{\mathrm{rx}}}\xspace}

\newcommand{\thetatx}{\ensuremath{\theta_{\mathrm{tx}}}\xspace}
\newcommand{\phitx}{\ensuremath{\phi_{\mathrm{tx}}}\xspace}

\newcommand{\thetarx}{\ensuremath{\theta_{\mathrm{rx}}}\xspace}
\newcommand{\phirx}{\ensuremath{\phi_{\mathrm{rx}}}\xspace}

\newcommand{\Ntx}{\ensuremath{N_{\mathrm{tx}}}\xspace}
\newcommand{\Nrx}{\ensuremath{N_{\mathrm{rx}}}\xspace}

\newcommand{\Mtx}{\ensuremath{M_{\mathrm{tx}}}\xspace}
\newcommand{\Mrx}{\ensuremath{M_{\mathrm{rx}}}\xspace}

\newcommand{\labelcl}{\mathrm{CL}}

\newcommand{\labelue}{\mathrm{UE}}
\newcommand{\labeltx}{\mathrm{tx}}
\newcommand{\labelrx}{\mathrm{rx}}

\newcommand{\labeltdd}{\mathrm{TDD}}
\newcommand{\labelfdd}{\mathrm{FDD}}

\newcommand{\snrtx}{\ensuremath{\snr_{\labeltx}}\xspace}
\newcommand{\snrrx}{\ensuremath{\snr_{\labelrx}}\xspace}

\newcommand{\sinrtx}{\ensuremath{\sinr_{\labeltx}}\xspace}
\newcommand{\sinrrx}{\ensuremath{\sinr_{\labelrx}}\xspace}

\newcommand{\inrtx}{\ensuremath{\inr_{\labeltx}}\xspace}
\newcommand{\inrrx}{\ensuremath{\inr_{\labelrx}}\xspace}
\newcommand{\inrrxthresh}{\ensuremath{\inr_{\labelrx}^{\mathrm{tgt}}}\xspace}

\newcommand{\inrrxmin}{\ensuremath{\inr_{\labelrx}^{\mathrm{min}}}\xspace}

\newcommand{\snrrxbar}{\ensuremath{\overline{\snr}_{\labelrx}}\xspace}

\newcommand{\capfd}{\ensuremath{C_{\mathrm{fd}}}\xspace}

\newcommand{\sefd}{\ensuremath{R_{\mathrm{fd}}}\xspace}

\newcommand{\setx}{\ensuremath{R_{\labeltx}}\xspace}
\newcommand{\serx}{\ensuremath{R_{\labelrx}}\xspace}

\newcommand{\setdd}{\ensuremath{R_{\labeltdd}}\xspace}
\newcommand{\sefdd}{\ensuremath{R_{\labelfdd}}\xspace}

\newcommand{\hcl}{h_{\labelcl}\xspace}

\newcommand{\vhtx}{\vh_{\labeltx}\xspace}
\newcommand{\vhrx}{\vh_{\labelrx}\xspace}

\newcommand{\powernoiseue}{\ensuremath{\powernoise^{\labelue}}\xspace}
\newcommand{\powernoisebs}{\ensuremath{\powernoise^{\mathrm{BS}}}\xspace}

\newcommand{\deltatheta}{\ensuremath{\delta\theta}\xspace}
\newcommand{\deltaphi}{\ensuremath{\delta\phi}\xspace}
\newcommand{\Deltatheta}{\ensuremath{\Delta\theta}\xspace}
\newcommand{\Deltaphi}{\ensuremath{\Delta\phi}\xspace}

\newcommand{\Deltavartheta}{\ensuremath{\Delta\vartheta}\xspace}
\newcommand{\Deltavarphi}{\ensuremath{\Delta\varphi}\xspace}

\newcommand{\setnbr}{\ensuremath{\mathcal{N}}\xspace}
\newcommand{\setnbrth}{\ensuremath{\mathcal{N}_{\theta}}\xspace}
\newcommand{\setnbrph}{\ensuremath{\mathcal{N}_{\phi}}\xspace}

\newcommand{\precbbar}{\ensuremath{\bar{\mathcal{F}}}\xspace}
\newcommand{\comcbbar}{\ensuremath{\bar{\mathcal{W}}}\xspace}

\newcommand{\sigmatx}{\ensuremath{\sigma_{\mathrm{tx}}}\xspace}
\newcommand{\sigmarx}{\ensuremath{\sigma_{\mathrm{rx}}}\xspace}
\newcommand{\sigmatxsq}{\ensuremath{\sigmatx^2}\xspace}
\newcommand{\sigmarxsq}{\ensuremath{\sigmarx^2}\xspace}

\newcommand{\bitsphase}{\ensuremath{{b_{\mathrm{phs}}}}\xspace}

\def\vf{{\vec{f}}}

\def\vh{{\vec{h}}}

\def\vw{{\vec{w}}}
\def\vx{{\vec{x}}}
\def\vy{{\vec{y}}}

\def\vone{{\vec{1}}}

\def\mA{{\mat{A}}}

\def\mF{{\mat{F}}}

\def\mH{{\mat{H}}}

\def\mW{{\mat{W}}}

\usepackage{xspace}
\usepackage[acronym,nogroupskip,nonumberlist,nopostdot]{glossaries}
\makeglossaries

\usepackage{xcolor}

\newacronym{snr}{SNR}{signal-to-noise ratio}
\newacronym{sinr}{SINR}{signal-to-interference-plus-noise ratio}
\newacronym{inr}{INR}{interference-to-noise ratio}
\newacronym{sir}{SIR}{signal-to-interference ratio}
\newacronym{sqr}{SQR}{signal-to-quantization-noise ratio}
\newacronym{sqnr}{SQNR}{signal-to-quantization-plus-noise ratio}
\newacronym{ian}{IAN}{interference as noise}
\newacronym{ber}{BER}{bit error rate}
\newacronym{pn}{PN}{pseudorandom noise}
\newacronym{bfsk}{BFSK}{binary frequency shift keying}
\newacronym{fh}{FH}{frequency-hopped}
\newacronym{fh-bfsk}{FH-BFSK}{frequency-hopped binary frequency shift keying}
\newacronym{crc}{CRC}{cyclic redundancy check}
\newacronym{isi}{ISI}{intersymbol interference}
\newacronym{dsss}{DSSS}{direct-sequence spread spectrum}
\newacronym{ofdm}{OFDM}{orthogonal frequency-division multiplexing}
\newacronym{ofdma}{OFDMA}{orthogonal frequency-division multiple access}
\newacronym{sdr}{SDR}{software-defined radio}
\newacronym{tx}{TX}{transmitter}
\newacronym{rx}{RX}{receiver}
\newacronym{fdd}{FDD}{frequency-division duplexing}
\newacronym{tdd}{TDD}{time-division duplexing}
\newacronym{xdd}{XDD}{cross-division duplexing}
\newacronym{fdma}{FDMA}{frequency-division multiple access}
\newacronym{tdma}{TDMA}{time-division multiple access}
\newacronym{sdma}{SDMA}{space-division multiple access}
\newacronym[plural=MPCs]{mpc}{MPC}{multipath component}

\newacronym{ls}{LS}{least-squares}
\newacronym{lms}{LMS}{least mean squares}
\newacronym{rls}{RLS}{recursive least-squares}
\newacronym{rzf}{RZF}{regularized zero-forcing}
\newacronym{mmse}{MMSE}{minimum mean square error}
\newacronym{lmmse}{LMMSE}{linear minimum mean square error}
\newacronym{mse}{MSE}{mean square error}
\newacronym{fft}{FFT}{fast Fourier transform}
\newacronym{dft}{DFT}{discrete Fourier transform}
\newacronym{dtft}{DTFT}{discrete-time Fourier transform}
\newacronym{ctft}{CTFT}{continuous-time Fourier transform}
\newacronym{ml}{ML}{machine learning}
\newacronym[plural=NNs]{nn}{NN}{neural network}
\newacronym[plural=RNNs]{rnn}{RNN}{recurrent neural network}
\newacronym[plural=ADCs]{adc}{ADC}{analog-to-digital converter}
\newacronym[plural=DACs]{dac}{DAC}{digital-to-analog converter}
\newacronym[plural=FPGAs]{fpga}{FPGA}{field-programmable gate array}
\newacronym{evm}{EVM}{error vector magnitude}
\newacronym{enob}{ENOB}{effective number of bits}
\newacronym{zf}{ZF}{zero-forcing}
\newacronym{rv}{r.v.}{random variable}
\newacronym{omp}{OMP}{orthogonal matching pursuit}
\newacronym{svd}{SVD}{singular value decomposition}
\newacronym{fir}{FIR}{finite impulse response}

\newacronym{agc}{AGC}{automatic gain control}
\newacronym{rf}{RF}{radio frequency}
\newacronym{if}{IF}{intermediate frequency}
\newacronym{los}{LOS}{line-of-sight}
\newacronym{nlos}{NLOS}{non-line-of-sight}
\newacronym{ple}{PLE}{path loss exponent}
\newacronym[plural=dB,firstplural=decibels (dB)]{db}{dB}{decibel}
\newacronym[plural=dBm,firstplural=decibel milliwatts (dBm)]{dbm}{dBm}{decibel milliwatts}
\newacronym{pa}{PA}{power amplifier}
\newacronym{lna}{LNA}{low noise amplifier}
\newacronym{vga}{VGA}{variable-gain amplifier}
\newacronym{cw}{CW}{continuous wave}
\newacronym{papr}{PAPR}{peak-to-average power ratio}
\newacronym{usrp}{USRP}{Universal Software Radio Peripheral}
\newacronym{irr}{IRR}{image rejection ratio}
\newacronym{lo}{LO}{local oscillator}
\newacronym{vm}{VM}{vector modulator}
\newacronym{mmwave}{mmWave}{millimeter-wave}
\newacronym{thz}{THz}{terahertz}
\newacronym{ris}{RIS}{reconfigurable intelligent surface}
\newacronym{eirp}{EIRP}{effective isotropic radiated power}
\newacronym{rsrp}{RSRP}{reference signal received power}

\newacronym{csma}{CSMA}{carrier-sense multiple access}
\newacronym{csmaca}{CSMA/CA}{carrier-sense multiple access with collision avoidance}
\newacronym{csmacd}{CSMA/CD}{carrier-sense multiple access with collision detection}
\newacronym{mac}{MAC}{medium access control}
\newacronym{phy}{PHY}{physical layer}
\newacronym{3gpp}{3GPP}{3rd Generation Partnership Project}
\newacronym{4g}{4G}{fourth generation}
\newacronym{lte}{LTE}{Long-Term Evolution}
\newacronym{4glte}{4G LTE}{\gls{4g} \gls{lte}}
\newacronym{5g}{5G}{fifth generation}
\newacronym{nr}{NR}{New Radio}
\newacronym{5gnr}{5G NR}{5G New Radio}
\newacronym{ieee}{IEEE}{Institute of Electrical and Electronics Engineers}
\newacronym{wifi}{Wi-Fi}{IEEE 802.11}
\newacronym{lan}{LAN}{local area network}
\newacronym{wlan}{WLAN}{wireless local area network}
\newacronym{bs}{BS}{base station}
\newacronym{sbs}{SBS}{small-cell base station}
\newacronym{mbs}{MBS}{macrocell base station}
\newacronym{ue}{UE}{user equipment}
\newacronym{ul}{UL}{uplink}
\newacronym{dl}{DL}{downlink}
\newacronym{qos}{QoS}{Quality of Service}
\newacronym{fcc}{FCC}{Federal Communications Commission}
\newacronym{iab}{IAB}{integrated access and backhaul}
\newacronym{fab}{FAB}{fixed access and backhaul}
\newacronym{fwa}{FWA}{fixed wireless access}
\newacronym{hetnet}{HetNet}{heterogeneous network}

\newacronym{siso}{SISO}{single-input single-output}
\newacronym{mimo}{MIMO}{multiple-input multiple-output}
\newacronym{sumimo}{SU-MIMO}{single-user \gls{mimo}}
\newacronym{mumimo}{MU-MIMO}{multi-user \gls{mimo}}
\newacronym{bf}{BF}{beamforming}
\newacronym{ca}{CA}{constant amplitude}
\newacronym{ula}{ULA}{uniform linear array}
\newacronym{upa}{UPA}{uniform planar array}
\newacronym[\glslongpluralkey={angles of arrival}]{aoa}{AoA}{angle of arrival}
\newacronym[\glslongpluralkey={angles of departure}]{aod}{AoD}{angle of departure}
\newacronym{dof}{DoF}{degrees of freedom}
\newacronym{csi}{CSI}{channel state information}
\newacronym{csit}{CSIT}{\gls{csi} at the transmitter}
\newacronym{csir}{CSIR}{\gls{csi} at the receiver}
\newacronym{cs}{CS}{compressed sensing}

\newacronym{fd}{FD}{in-band full-duplex}
\newacronym{hd}{HD}{half-duplex}
\newacronym{si}{SI}{self-interference}
\newacronym{sic}{SIC}{self-interference cancellation}
\newacronym{soi}{SoI}{signal of interest}
\newacronym{asic}{A-SIC}{analog \acrlong{sic}}
\newacronym{dsic}{D-SIC}{digital \gls{sic}}
\newacronym{star}{STAR}{simultaneous transmit and receive}
\newacronym{warp}{WARP}{Wireless Open-Access Research Platform}
\newacronym{bfc}{BFC}{beamforming cancellation}
\newacronym{ipi}{IPI}{inter-panel-interference}
\newacronym{ipic}{IPIC}{inter-panel-interference cancellation}

\newacronym{pdf}{PDF}{probability density function}
\newacronym{cdf}{CDF}{cumulative density function}
\newacronym{iid}{i.i.d.}{independently and identically distributed}

\newacronym{elf}{ELF}{extremely low frequency}
\newacronym{slf}{SLF}{super low frequency}
\newacronym{ulf}{ULF}{ultra low frequency}
\newacronym{vlf}{VLF}{very low frequency}
\newacronym{lf}{LF}{low frequency}
\newacronym{mf}{MF}{medium frequency}
\newacronym{hf}{HF}{high frequency}
\newacronym{vhf}{VHF}{very high frequency}
\newacronym{uhf}{UHF}{ultra high frequency}
\newacronym{shf}{SHF}{super high frequency}
\newacronym{ehf}{EHF}{extremely high frequency}
\newacronym{thf}{THF}{tremendously high frequency}

\newcommand{\mmwave}{\gls{mmwave}\xspace}
\newcommand{\mimo}{\gls{mimo}\xspace}

\newcommand{\rf}{\gls{rf}\xspace}

\newcommand{\si}{\acrshort{si}\xspace}

\newcommand{\sic}{\gls{sic}\xspace}

\newcommand{\iab}{\gls{iab}\xspace}

\newcommand{\lna}{\gls{lna}\xspace}
\newcommand{\pa}{\gls{pa}\xspace}
\newcommand{\lnas}{\glspl{lna}\xspace}
\newcommand{\pas}{\glspl{pa}\xspace}

\newcommand{\adc}{\gls{adc}\xspace}
\newcommand{\adcs}{\glspl{adc}\xspace}

\newcommand{\gcdf}{\gls{cdf}\xspace}

\newcommand{\bs}{\acrlong{bs}\xspace}
\newcommand{\bss}{\acrlongpl{bs}\xspace}
\newcommand{\gsnr}{\gls{snr}\xspace}
\newcommand{\ginr}{\gls{inr}\xspace}
\newcommand{\gsinr}{\gls{sinr}\xspace}

\newcommand{\gpsnr}{\glspl{snr}\xspace}
\newcommand{\gpinr}{\glspl{inr}\xspace}
\newcommand{\gpsinr}{\glspl{sinr}\xspace}

\newcommand{\tdd}{\gls{tdd}\xspace}
\newcommand{\fdd}{\gls{fdd}\xspace}

\newcommand{\dacs}{\glspl{dac}\xspace}

\newcommand{\subsecref}[1]{Subsection~\ref{#1}}

\newcommand{\figref}[1]{\figurename~\ref{#1}}

\newcommand{\steer}{\textsc{Steer}\xspace}

\usepackage{newtxtext}       %
\usepackage[varvw]{newtxmath}       

\begin{document}

\title*{Full-Duplex Transceivers for Next-Generation Wireless Communication Systems}
\author{Ian P.~Roberts and Himal A.~Suraweera}
\institute{Ian P.~Roberts \at Wireless Networking and Communications Group, Department of Electrical and Computer Engineering, University of Texas at Austin, Austin, Texas 78712, USA, \email{ipr@utexas.edu}
\and Himal A.~Suraweera \at Department of Electrical and Electronic Engineering, University of Peradeniya, Peradeniya 20400, Sri Lanka, \email{himal@eng.pdn.ac.lk}}
%
%

\maketitle

\abstract{%
Wireless communication systems can be enhanced at the link level, in medium access, and at the network level when transceivers are equipped with full-duplex capability: the transformative ability to simultaneously transmit and receive over the same frequency spectrum. 
Effective methods to cancel self-interference are required to facilitate full-duplex operation, which we overview herein in the context of traditional radios, along with those in next-generation wireless networks.
We highlight advances in self-interference cancellation that leverage machine learning, and we summarize key considerations and recent progress in full-duplex millimeter-wave systems and their application in integrated access and backhaul.
We present example design problems and noteworthy findings from recent experimental research to introduce and motivate the advancement of full-duplex millimeter-wave systems.
We conclude this chapter by forecasting the future of full-duplex and outlining important research directions that warrant further study.
}

\abstract*{Wireless communication systems can be enhanced at the link level, in medium access, and at the network level when transceivers are equipped with full-duplex capability: the transformative ability to simultaneously transmit and receive over the same frequency spectrum. Effective methods to cancel self-interference are required to facilitate full-duplex operation, which we overview herein in the context of traditional radios, along with those in next-generation wireless networks. We highlight advances in self-interference cancellation that leverage machine learning, and we summarize key considerations and recent progress in full-duplex millimeter-wave systems and their application in integrated access and backhaul. We present example design problems and noteworthy findings from recent experimental research to introduce and motivate the advancement of full-duplex millimeter-wave systems. We conclude this chapter by forecasting the future of full-duplex and outlining important research directions that warrant further study.}



\glsresetall

\section{Introduction} \label{sec:introduction}
For more than a century, wireless communication systems have almost exclusively operated in a half-duplex fashion, where transmission and reception of radio waves have typically been separated---or orthogonalized---in the time domain, frequency domain, or both.
Put simply, signals transmitted or received by a traditional half-duplex system exist in different frequency bands or at different times, referred to as \fdd and \tdd, respectively.
Half-duplex operation is necessitated by the manifestation of \textit{self-interference} (SI) when a transceiver attempts to receive signals while simultaneously transmitting in the same spectrum.
In most cases, \si is many orders of magnitude stronger than a relatively weak signal-of-interest (or \textit{desired receive signal}), which has presumably propagated tens or hundreds of meters.
This makes it virtually impossible to accurately recover the desired receive signal from their combination without taking additional measures to mitigate the effects of \si \cite{ashu_inband_jsac_2014,kolodziej_techniques_2019}. 
By receiving in neighboring frequency spectrum or on a separate time slot as its transmissions, a half-duplex transceiver can avoid inflicting \si onto a desired receive signal, hence the usage of \fdd and \tdd.

By their nature, \fdd and \tdd both consume radio resources by dedicating time-frequency resources to either transmission or reception.
Of course, this would not be an issue if practical systems were not resource-constrained.
In reality, all practical wireless communication systems operate on limited time-frequency resources. 
At the very least, most systems are confined to certain frequency spectrum by regulatory bodies, such as the \gls{fcc} in the United States.
The consumption of radio resources by half-duplex operation has motivated researchers to explore in-band \textit{full-duplex} operation\footnote{We use the term ``full-duplex'' to refer to in-band full-duplex operation, in particular, as opposed to out-of-band full-duplex, which has been used to describe systems capable of simultaneously transmitting and receiving via \fdd.} \cite{division_free_1998,stanford_achieving,ashu_inband_jsac_2014,hong_sicapps_2014,himal_fd_book_2020,cbc_mag_2015}.
Starting in the late 2000s, researchers began heavily investigating and developing means to mitigate \si and bring full-duplex to life.
Since then, full-duplex has matured and has recently found new life in \mmwave networks \cite{xia_2017,roberts_wcm}, in joint communication and sensing \cite{jcas_wcm_2021}, and through advancements in machine learning \cite{stimming_implement_2018}.

In this chapter, we introduce full-duplex operation and highlight its enhancements.
Then, we overview full-duplex solutions for traditional radios and those for modern and next-generation wireless systems.
We conclude with a look ahead at the future of full-duplex to motivate and steer its research, development, and deployment.

\begin{figure}[t]
    \centering
    \includegraphics[width=0.29\linewidth,height=\textheight,keepaspectratio]{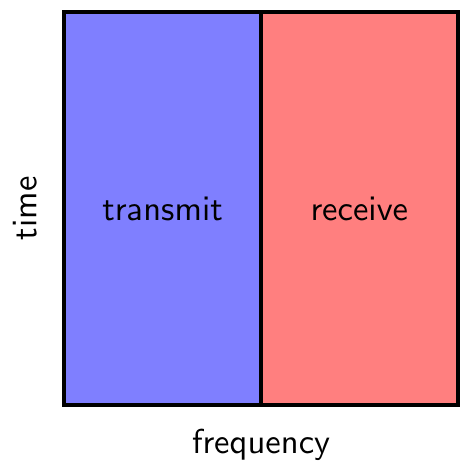}
    \qquad
    \includegraphics[width=0.29\linewidth,height=\textheight,keepaspectratio]{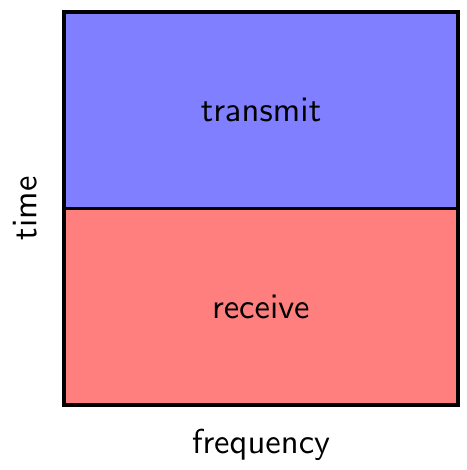}
    \qquad
    \includegraphics[width=0.29\linewidth,height=\textheight,keepaspectratio]{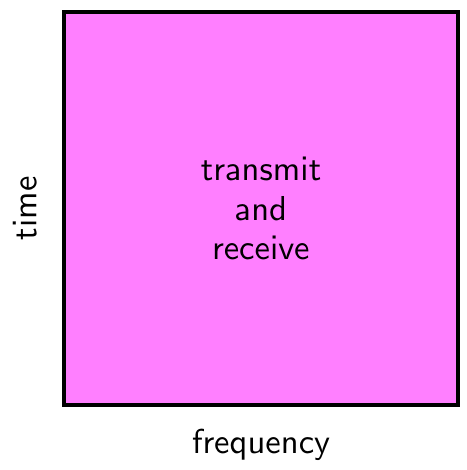}
    \caption{The time-frequency resources consumed by transmission and reception with (left) \fdd; (middle) \tdd; and (right) full-duplex. In practice, guard bands are usually necessary for both \fdd and \tdd, which consumes additional radio resources.}
    \label{fig:resources}
\end{figure}



\subsection{What is Full-Duplex?}
Full-duplex capability allows a transceiver to concurrently transmit and receive over the same frequency spectrum.
In other words, transmission and reception can both make use of the full available frequency spectrum all the time.
As mentioned, when operating in this full-duplex fashion, \si is inflicted onto a desired receive signal.
To equip communication systems with full-duplex capability, engineers have developed a variety of ways to mitigate \si using \rf components, analog and digital filters, and other creative means.
We will outline a variety of these strategies in this chapter.
If \si can be sufficiently mitigated, a full-duplex transceiver can reliably receive while transmitting in-band, unlocking a number of enhancements, which we overview in the next section.

\begin{figure}[t]
    \centering
    \includegraphics[width=\linewidth,height=0.15\textheight,keepaspectratio]{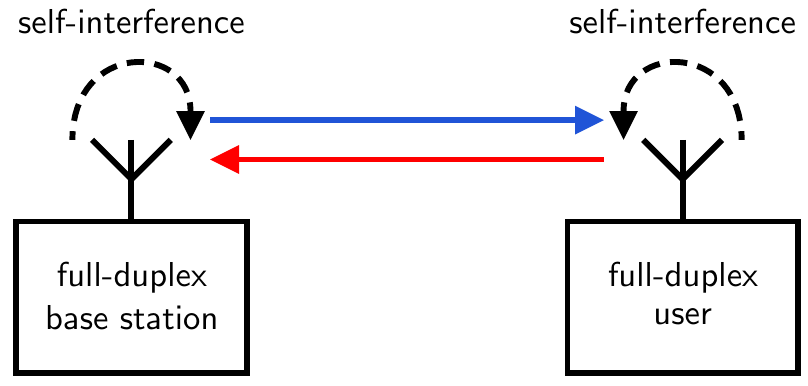}
    \caption{A full-duplex base station transmitting and receiving in-band with a full-duplex user. \si manifests at both devices, requiring each to take measures to sufficiently cancel it.}
    \label{fig:fdx-a}
\end{figure}

As depicted in \figref{fig:fdx-a}, a full-duplex base station may transmit and receive concurrently with a neighboring user that also has full-duplex capability.
This makes better use of radio resources, and as intuition may suggest, leads to a potential doubling of spectral efficiency, as compared to half-duplex techniques.
In other words, radio resources are being used twice as efficiently with full-duplex, since they are used for both transmission and reception, rather than divided as with \fdd and \tdd, as illustrated in \figref{fig:resources}.

\begin{figure}[t]
    \centering
    \includegraphics[width=\linewidth,height=0.18\textheight,keepaspectratio]{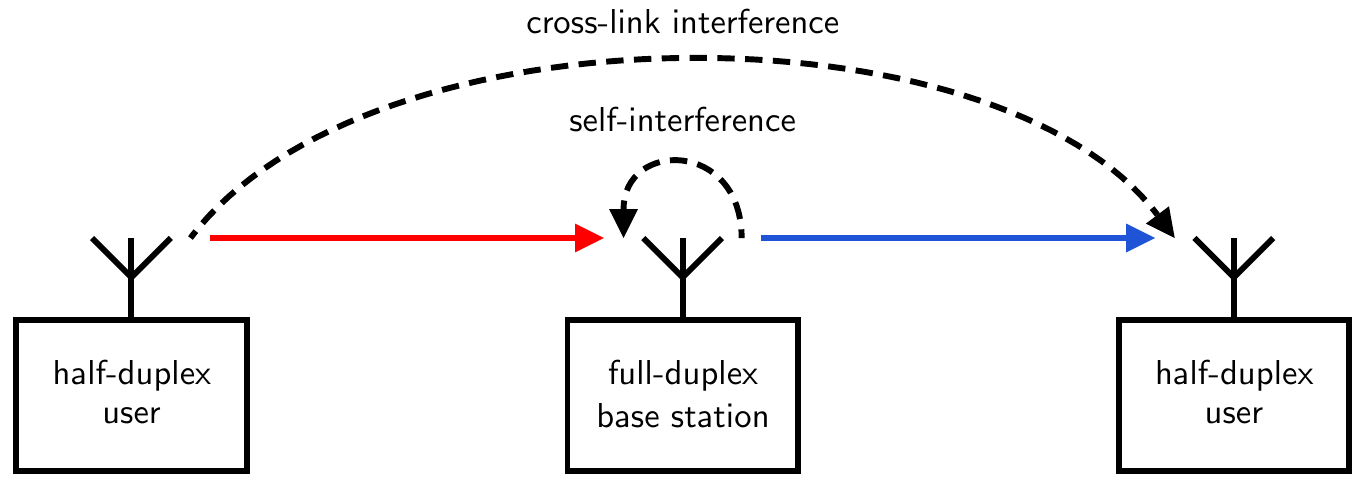}
    \caption{A full-duplex base station transmitting to one half-duplex user while receiving from another half-duplex user in-band. \si manifests at the base station, whereas cross-link interference is inflicted onto the user receiving by the user transmitting.}
    \label{fig:fdx-b}
\end{figure}

\figref{fig:fdx-b} depicts another full-duplex operating mode, where a full-duplex base station transmits to a half-duplex user while receiving from another half-duplex user.
This mode of operation can also potentially double spectral efficiency.
It is important to note that cross-link interference is inflicted on the user receiving by the user transmitting, the level of which depends on a number of factors, chiefly the users' locations and the environment.

\begin{svgraybox}
    \textbf{Takeaways.}
    Full-duplex operation is an exciting alternative to half-duplexing with \fdd and \tdd since it makes better use of radio resources. To enable full-duplex operation, however, self-interference must be mitigated sufficiently in order to reliably recover desired receive signals while transmitting in-band.
\end{svgraybox}

\section{Enhancements Introduced by Full-Duplex}
Compared to half-duplex operation, full-duplex can introduce enhancements to communication systems on a link level \cite{himal_fd_book_2020,ashu_inband_jsac_2014,kolodziej_techniques_2019,hong_sicapps_2014}, in medium access and spectrum sharing \cite{liao_protocol_2015,liao_cognitive_2014,liao_mag_2015,song_csma_2016,thilina_mag_2015,mishra_collision}, and at the network level \cite{ozgur_isit_2017,gupta_fdiab}---which we overview in this section.

\subsection{Link-Level Analysis}
The impacts of full-duplex can be readily observed by examining familiar link-level expressions foundational to wireless communication systems.
To do so, consider the full-duplex operating mode depicted in \figref{fig:fdx-b}, where a full-duplex base station communicates with two half-duplex users.
Taking the perspective of the full-duplex base station, we refer to its \textit{transmit link} and \textit{receive link}.
The full-duplex capacity of the system, in the absence of interference, can be written as
\begin{align}
\capfd = \logtwo{1+\snrtx} + \logtwo{1+\snrrx},
\end{align}
where $\snrtx$ and $\snrrx$ are the maximum \gpsnr of the transmit and receive links, respectively.
If employing \tdd to duplex transmission and reception instead of full-duplex, the achievable sum spectral efficiency is
\begin{align}
\setdd = \alpha \cdot \logtwo{1+\snrtx} + \parens{1-\alpha} \cdot \logtwo{1+\snrrx},
\end{align}
where $\alpha$ is the fraction of time allocated to transmission, with the remainder allocated to reception.
If employing \fdd, this achievable sum spectral efficiency becomes
\begin{align}
\sefdd = \alpha \cdot \logtwo{1+\frac{\snrtx}{\alpha}} + (1-\alpha) \cdot \logtwo{1+\frac{\snrrx}{1-\alpha}},
\end{align}
where $\alpha$ is the fraction of the total bandwidth allocated to transmission, with the remainder allocated to reception.
Here, since the integrated noise power is proportional to bandwidth, \fdd enjoys \gsnr increases inversely proportional to bandwidth.
The expression of $\setdd$ as presented implicitly assumes an instantaneous transmit power constraint.
Under an average transmit power constraint, which is less practical, $\setdd$ and $\sefdd$ coincide \cite{heath_lozano}.

Under full-duplex operation, the \gls{sinr} of a desired receive signal (i.e., a signal-of-interest) on a given link is
\begin{align}
\sinr 
= \frac{\powerdes}{\powernoise + \powerint}
= \frac{\snr}{1 + \inr}, \label{eq:sinr}
\end{align}
where $\powerdes$ is the power of the desired receive signal; $\powernoise$ is the noise power; and $\powerint$ is the power of interference (i.e., \si on the receive link, cross-link interference on the transmit link).
The right-hand side of \eqref{eq:sinr} can be obtained by dividing the numerator and denominator by the noise power $\powernoise$, where $\snr$ is the \gls{snr} of the desired receive signal and $\inr$ is the \gls{inr}.
The power of \si---and hence \inr on the receive link---depends on the quality of \si mitigation employed at the full-duplex transceiver.
For now, we can consider the degree of (residual) \si $\powersi$ as being some level below the transmit power $\powertx$ at the full-duplex transceiver.
For instance, suppose our full-duplex base station is capable of reducing \si to a power level of
\begin{align}
\powersi = \frac{\powertx}{L},
\end{align}
where $L$ is the total amount of mitigation (i.e., cancellation) achieved by its full-duplex solution.
The total amount of \si mitigation $L$ may capture a variety of efforts, including antenna isolation and/or additional \si cancellation filtering, which we will discuss further in the next section.
Since the power of a desired receive signal $\powerdes$ can be quite close to the noise floor $\powernoise$ in practical systems, \si power $\powersi$ must be near or below the noise floor to ensure it does not prohibitively erode full-duplex resource gains. 
This means is not uncommon for $L$ to be on the order of $100$ dB for practical systems.
Consider a transmit power of $\powertx = 20$ dBm and a noise floor of $\powernoise = -90$ dBm: $L = 110$ dB is required for $\powersi = \powernoise$ (i.e., $\inr = 0$ dB).
Achieving this degree of mitigation is precisely what has hindered the adoption of full-duplex since the advent of wireless communications and what has made successful demonstrations of full-duplex so impressive \cite{cbc_mag_2015,duarte_experiment_2012,stanford_practical,stanford_full_duplex_radios,korpi_dissertation}.



\begin{svgraybox}
    \textbf{A summary of key power ratios of the full-duplex system in \figref{fig:fdx-b}.}
    \begin{description}[$\sinrrx$]
        \item[$\snrtx$]{Strength of the desired signal on the transmit link.}
        \item[$\snrrx$]{Strength of the desired signal on the receive link.}
        \item[$\inrtx$]{Strength of cross-link interference on the transmit link.}
        \item[$\inrrx$]{Strength of \si on the receive link.}
        \item[$\sinrtx$]{Effective quality of the desired signal on the transmit link.}
        \item[$\sinrrx$]{Effective quality of the desired signal on the receive link.}
    \end{description}
\end{svgraybox}

When full-duplexing transmission and reception, the maximum achievable spectral efficiency when treating residual interference as noise can be expressed as
\begin{align}
\sefd = \logtwo{1+\sinrtx} + \logtwo{1+\sinrrx} \leq \capfd.
\end{align}
When $\sinrtx \to \snrtx$ and $\sinrrx \to \snrrx$, then $\sefd \to \capfd$.
These expressions illustrate the potential resource gains offered by full-duplex compared to \fdd and \tdd, since there are no pre-log fractions; the effectiveness of such depends heavily on the presence of low \si and low cross-link interference, however.

\begin{figure}[t]
    \centering
    \includegraphics[width=0.6\linewidth,height=\textheight,keepaspectratio]{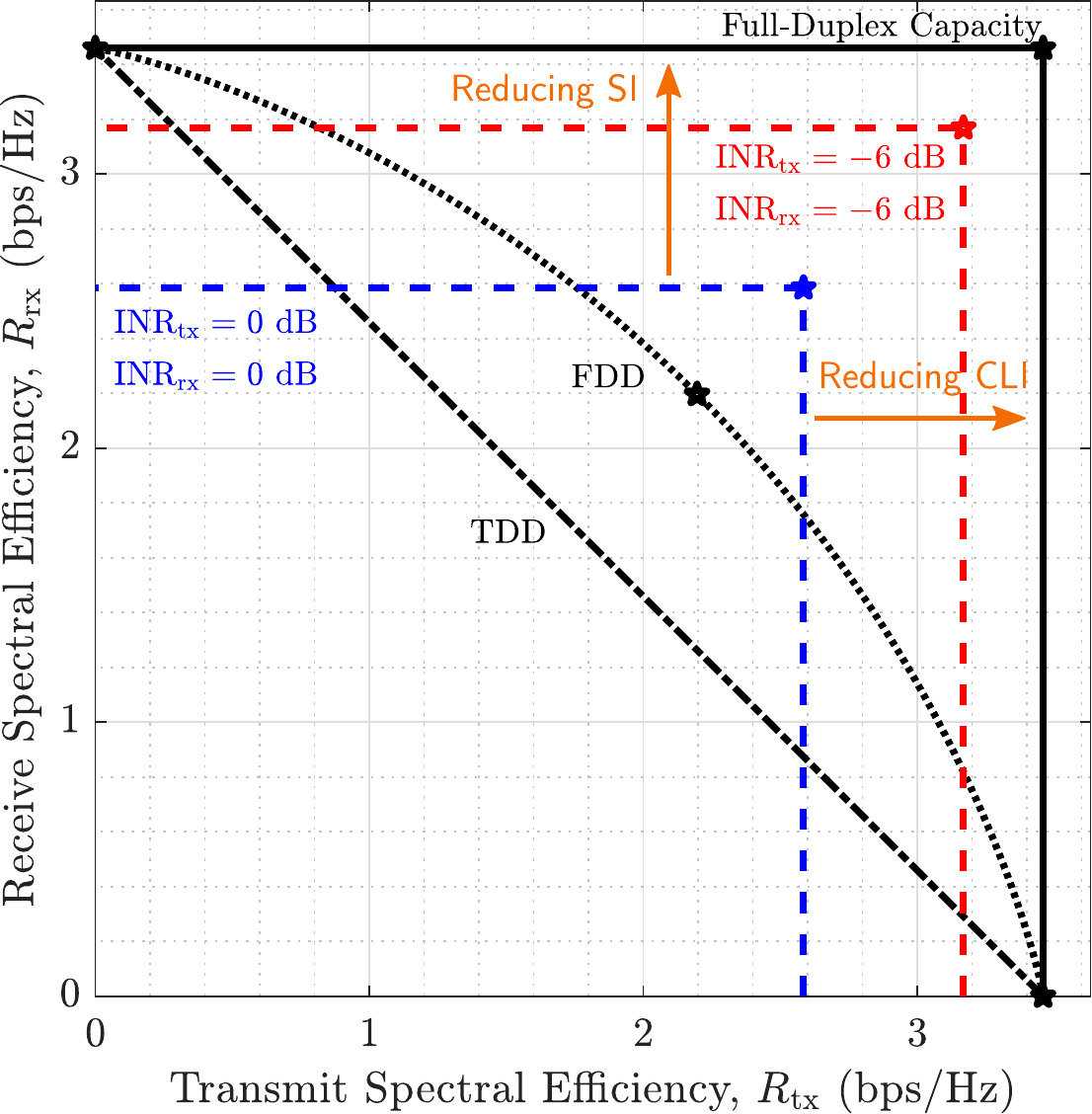}
    \caption{The rate region boundaries for various multiplexing strategies when $\snrtx = \snrrx = 10$ dB. Full-duplex, shown as dashed blue and red lines, can outperform \tdd and \fdd with \si and cross-link interference (shown as CLI) sufficiently mitigated. Stars ($\star$) indicate points that maximize the sum spectral efficiency.}
    \label{fig:region}
\end{figure}

In \figref{fig:region}, we compare full-duplex operation against half-duplex operation with \fdd and \tdd by plotting their \textit{rate regions} for various levels of \si $\inrrx$ and cross-link interference, where $\snrtx = \snrrx = 10$ dB. 
Here, each line depicts the boundary of its rate region, encompassing all feasible transmit-receive spectral efficiency pairs $\parens{\setx,\serx}$, and each star ($\star$) indicates the point that maximizes the sum spectral efficiency.
The simple time-sharing nature of \tdd is shown as the diagonal line connecting the two points of greedy time-sharing.
\fdd offers improvements over \tdd, courtesy of its \gsnr gains when shrinking bandwidth, as mentioned.
When \si power is mitigated to a level equal to the noise power (i.e., $\inrrx = 0$ dB) and cross-link interference is also at the noise floor, the sum spectral efficiency can clearly exceed \fdd and \tdd, but only marginally.
As \si and cross-link interference are reduced to below the noise floor, however, the achievable rate region of full-duplex expands, approaching that of its capacity region.

\begin{svgraybox}
    \textbf{At lower \gpsnr}, lower \gpinr are required for appreciable full-duplex operation, since the effects of interference magnify due to the steep nature of $\logtwo{1+x}$ at low $x$, tightening the requirements \si and cross-link interference mitigation. In addition, the \gsnr gains introduced by \fdd are magnified since doubling $x$ nearly doubles $\logtwo{1+x} \approx x$ at low $x$, reducing the gap between \fdd and the full-duplex capacity.
\end{svgraybox}


\begin{svgraybox}
    \textbf{At higher \gpsnr}, the gap between full-duplex and half-duplex grows, and the effects of interference diminish due to the saturating nature of $\logtwo{1+x}$ at high $x$. Higher \gpinr can be tolerated at high \gpsnr, relaxing the requirements on mitigating \si and cross-link interference.
\end{svgraybox}


%

%
%
%
%
%
%
%
%
%



\subsection{In Medium Access and Spectrum Sharing}
Half-duplex transceivers have been ubiquitous in wireless networks, and for good reason, communication standards and practices have been built on this half-duplex assumption.
The ability to transmit and receive simultaneously and in-band is a transformative technology that can unlock new approaches to medium access and spectrum sharing that are more efficient than those built on a half-duplex assumption.
In turn, full-duplex can facilitate wireless networks that deliver higher throughput, inflict lower interference, and make better use of spectrum.


\begin{figure}[t]
    \centering
    \includegraphics[width=\linewidth,height=0.15\textheight,keepaspectratio]{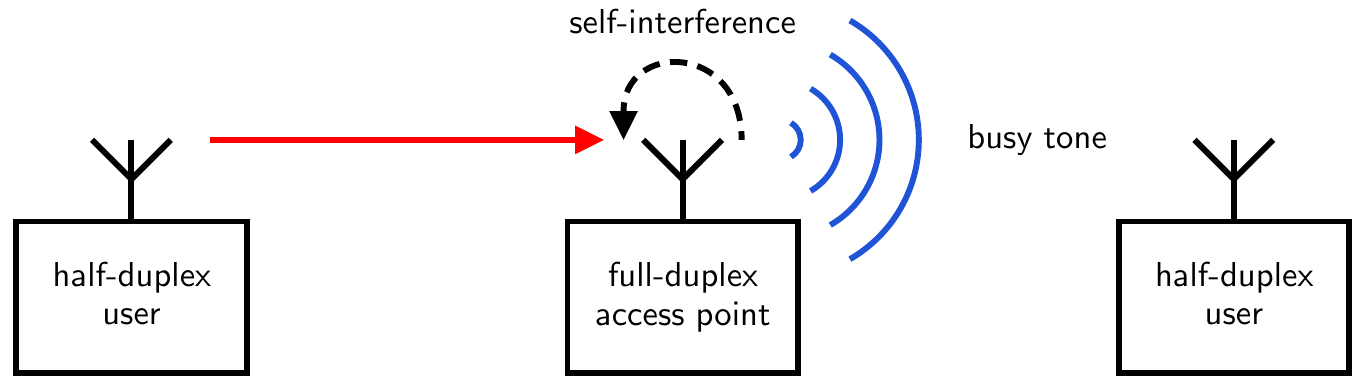}
    \caption{While receiving from one user, a full-duplex access point broadcasts a busy tone instructing all other users not to transmit, preventing collisions and overcoming the hidden node problem.}
    \label{fig:mac}
\end{figure}

\runinhead{Overcoming the Hidden Node Problem.}
To illustrate this, consider the famous hidden node problem in wireless networks: 
if two users are distant from one another but both within earshot of a nearby access point, the two users may be unaware when the other is transmitting to the access point.
This can lead to collisions at the access point---and hence a waste of radio resources---if not dealt with.
Conventional approaches to overcome this use handshaking between users and the access point to grant a user permission to transmit (e.g., request-to-send (RTS) and clear-to-send (CTS) mechanisms) along with random backoffs.
By upgrading the access point with full-duplex capability, more efficient approaches to medium access can be employed \cite{liao_protocol_2015,song_csma_2016,thilina_mag_2015}.
For instance, as illustrated in \figref{fig:mac}, the full-duplex access point can broadcast a busy tone anytime it is receiving from a user.
This busy tone can be heard by all users in the network, informing them to not transmit.
Compared to conventional approaches to medium access, this approach consumes no additional radio resources and avoids the overhead associated with handshaking between users and the access point and reduces latency.

\runinhead{Dynamic Spectrum Access and Cognitive Radios.}
The number of devices with wireless connectivity has grown profoundly over the past two decades and will seemingly continue for years to come.
The amount of available spectrum has not grown at a comparable pace, however.
In light of this, researchers have proposed spectrum sharing and cognitive radios to dynamically and opportunistically make better use of frequency spectrum when it is free \cite{haykin_cognitive_2005}.
For instance, a cognitive radio may sense a frequency band to see if it is being used by nearby devices.
If there appears to be no activity, the cognitive radio may begin transmitting information.
Periodically, the cognitive radio may halt transmission to sense the channel to ensure that it does not inflict interference on incumbents that have rights to the band---a waste of resources if none are detected.
Full-duplex can make this process more efficient by empowering the cognitive radio to continuously sense the channel while transmitting \cite{liao_cognitive_2014,liao_mag_2015,thilina_mag_2015}.
This allows it to more efficiently transmit information, since it does not have to halt transmission to sense the channel, and enables the cognitive radio to react more quickly to the presence of incumbents.
Without full-duplex capability, the cognitive radio would presumably be unaware of the presence of incumbents until the end of its transmission, potentially causing interference that leads to collisions.
Techniques discussed for overcoming the hidden node problem can be applied in this setting, as well, to instruct nearby cognitive radios to not transmit.

\runinhead{Channel Sensing to Reduce the Effects of Interference.}
As a final example of full-duplex applied to medium access, we consider the case where an access point serves users in the presence of neighboring nodes that may inflict interference, as explored in \cite{mishra_collision}.
For instance, one can consider two Wi-Fi access points operating independently but near one another.
When one access point transmits to a user, successful reception at that user may be corrupted by neighboring interference, requiring the access point to retransmit the data.
Naturally, this consumes radio resources and can lead to delays in communication.
With full-duplex capability, the access point may sense the channel while transmitting, allowing it to potentially halt transmission to avoid collisions caused by neighboring interference and subsequently redirect resources to another user that may not be impacted by this interference \cite{mishra_collision,thilina_mag_2015}.
It is important to note that the requirements on mitigating \si are stricter when decoding data carried by a desired receive signal, compared to those for channel sensing, which is often merely detecting energy levels in a particular frequency band.

\begin{figure}[t]
    \centering
    \includegraphics[width=\linewidth,height=\textheight,keepaspectratio]{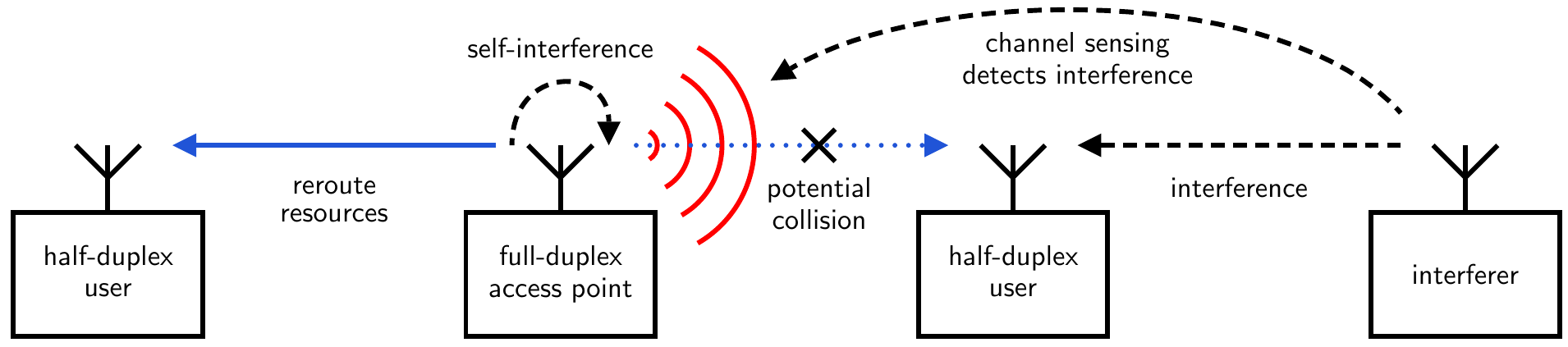}
    \caption{An access point equipped with full-duplex capability can sense the channel while transmitting, allowing it to reroute resources in the presence of interference that would otherwise cause collisions \cite{mishra_collision}.}
    \label{fig:detect}
\end{figure}



\subsection{Network-Level Enhancements}
Upgrading transceivers with full-duplex capability can be felt at the network level in a few ways.
Even when only some devices are equipped with full-duplex---while the remainder are half-duplex-capable---a wireless network can enjoy improvements in throughput and latency \cite{gupta_fdiab,ozgur_isit_2017}.
In fact, network throughput can magnify beyond the doubling of spectral efficiency we are familiar with at the link level with full-duplex \cite{gupta_fdiab,ozgur_isit_2017}.
This can be attributed to the fact that full-duplex can reduce multiplexing delays, reduce overhead associated with medium access control, and enable new ways to manage interference---all of which can improve network throughput.
We elaborate more on network-level enhancements of full-duplex in \subsecref{subsec:iab}. 

\begin{svgraybox}
    \textbf{Other applications.} There are applications of full-duplex technology beyond what was highlighted herein, such as in joint communication and sensing \cite{jcas_wcm_2021,xiao_jsac_2022}, secure communications \cite{wang_coml_2022}, military communications \cite{zheng_jamming_2013,riihonen_tactical_2017}, radar \cite{jcas_wcm_2021}, and more \cite{himal_fd_book_2020,hong_sicapps_2014}.
\end{svgraybox}

\section{Self-Interference Cancellation} \label{sec:sic}

\begin{figure}[t]
    \centering
    \includegraphics[width=\linewidth,height=\textheight,keepaspectratio]{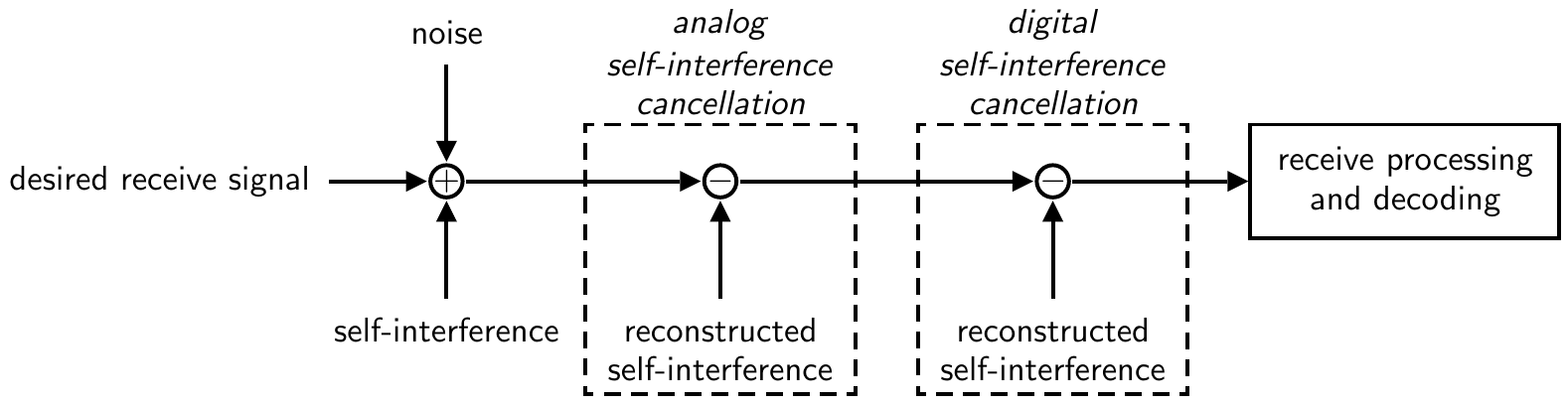}
    \caption{A received signal undergoes analog and digital SIC before undergoing conventional receive processing to recover desired receive data.}
    \label{fig:sic}
\end{figure}

Successfully equipping a device with full-duplex capability relies on mitigating---or \textit{cancelling}---\si to levels that are sufficiently low \cite{kolodziej_techniques_2019}.
The amount of \sic needed depends on the particular application.
In most settings, full-duplex solutions aim to cancel \si to near or below the receiver noise floor (i.e., roughly $\inrrx \leq 0$ dB).
This ensures that the full-duplex resource gains are not eroded by the presence of high \si, as highlighted in \figref{fig:region}.
The \textit{residual} \si is that which remains after efforts of \sic.
In this section, we outline methods of \sic in both the analog and digital domains.
Regardless of domain, the motivation behind \sic can largely be summarized as leveraging the fact that a transceiver has knowledge of its own transmitted signal and can therefore potentially reconstruct \si and subtract it from the received signal, leaving the desired portion virtually free of \si.
In many cases, a staged approach to \sic is employed as illustrated in \figref{fig:sic}, where a portion of \si is cancelled using an analog filter and a significant portion of the remainder is cancelled using digital filtering.
This staged approach is depicted in \figref{fig:asic-dsic}, where a full-duplex transceiver with separate antennas for transmission and reception employs an analog \sic filter between its antennas at \rf and a digital \sic filter.


\subsection{Analog Self-Interference Cancellation}
\begin{figure}[t]
    \centering
    \includegraphics[width=0.8\linewidth,height=\textheight,keepaspectratio]{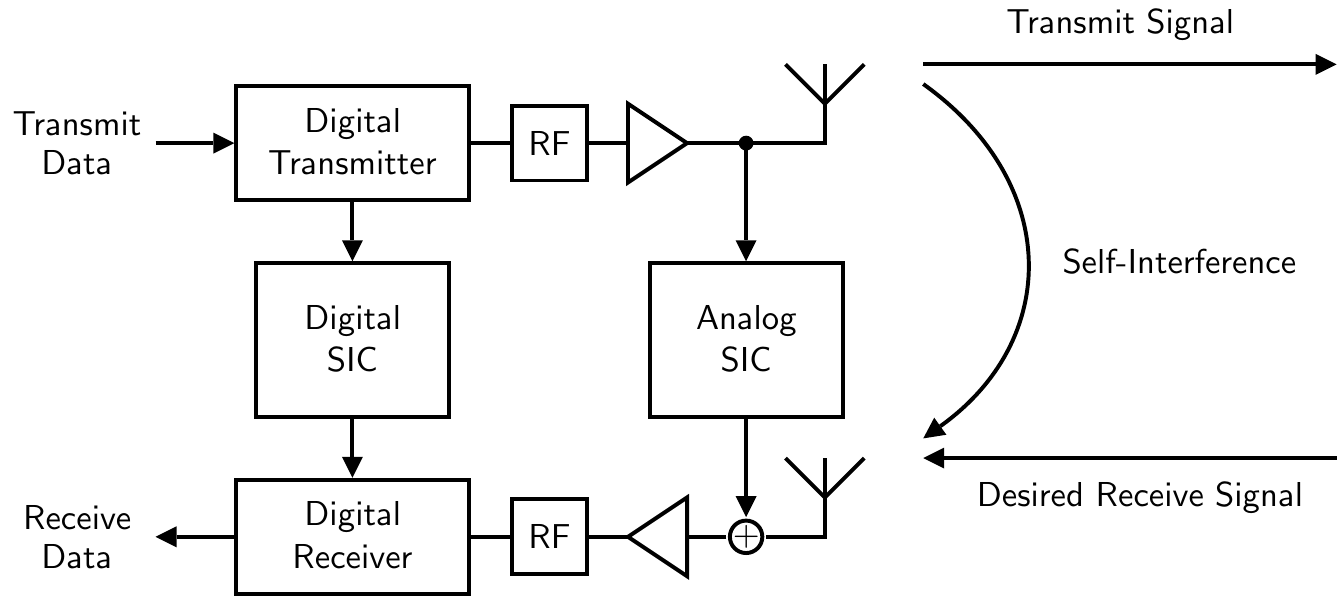}
    \caption{\si manifests when a full-duplex transceiver attempts to simultaneously transmit and receive using the same frequency spectrum. Here, separate antennas are used for transmission and reception, with analog and digital \sic used to reconstruct and subsequently cancel \si incurred at the receiver.}
    \label{fig:asic-dsic}
\end{figure}

As illustrated in \figref{fig:asic-dsic}, analog \sic typically exists as a digitally-controlled analog filter placed between the transmitter and receiver of a full-duplex transceiver.
Analog \sic filters come in many forms, existing as \rf, \acrlong{if}, and baseband circuitry, and even as optical filters \cite{kolodziej_techniques_2019,stanford_full_duplex_radios,suarez_optical_2009,han_optical_2017,su_optical_2020}.
Analog \sic is often driven by the tapping off a small portion of the upconverted \rf transmit signal.
This transmit signal undergoes filtering within the analog \sic filter before being injected at the receiver.
The injected signal is an inverted reconstruction of \si, which, when combined with the received signal, destructively combines with \si.
After this combining, there is some degree of residual \si due to imperfect reconstruction, which may stem from estimation errors, hardware limitations, and hardware imperfections.
By tapping off the transmit signal after the transmit chain, analog \sic will inherently incorporate transmit-side impairments unbeknownst to baseband, such as \pa nonlinearities, which have proven to be a dominant factor in \sic \cite{korpi_full-duplex_2014,korpi_dissertation}.
Other approaches, sometimes called digitally-assisted approaches, use a dedicated transmit chain to drive analog \sic, as opposed to tapping off the transmit signal directly \cite{kolodziej_techniques_2019}.
This approach cannot as well capture transmit-side impairments present in \si, however, since this second transmit chain naturally will not contain all artifacts of the true transmit chain.

\runinhead{Configuring an Analog Self-Interference Cancellation Filter.}
Tuning an analog \sic filter to effectively cancel \si largely consists of measuring \si and then configuring the filter to reconstruct its inverse. 
Analog \sic can be implemented as a time-domain filter or as a frequency-domain filter, meaning particular methods may vary but all tackle the same goal of reconstructing \si \cite{kolodziej_techniques_2019}.
One method of time-domain analog \sic is to estimate the impulse response of the \si channel and then configure the analog filter to produce this (inverted) impulse response estimate, effectively equalizing \si.
Estimation of the \si channel is typically executed by transmitting a pilot signal during a \textit{quiet period}, when no desired receive signal is present.
One difficulty with practically executing this method lay in the fact that estimation of the \si channel takes place digitally, meaning estimation of the channel of interest for analog \sic may be complicated by artifacts of the transmit and receive chains before and after analog \sic.
This can be further complicated by the fact that an analog filter may not have an ideal impulse response itself, making it difficult to reliably produce the desired impulse response.



As an attempt to overcome these challenges, another approach is to measure \si and then measure the impulse response of the analog filter.
Then, the filter can be configured to produce an inverted reconstruction of \si.
For instance, consider a column vector of measured \si time-domain samples $\vy$ (during a quiet period) and a matrix $\mA$ whose $i$-th column is the measured impulse response of the $i$-th tap of the filter.
Analog \gls{fir} filter weights $\vx$ can be computed to minimize the error in reconstructing an inverted copy of the measured \si as
\begin{align}
\vx\opt = \argmin_{\vx} \ \normtwo{-\vy - \mA\vx}^2,
\end{align}
which has the well-known closed form solution $\vx\opt = -\parens{\mA\ctrans\mA}\inv \mA\ctrans \vy$.
This approach has shown to be fairly robust, since it accounts for the imperfect impulse response of each of the filter's taps.

\begin{figure}[t]
    \centering
    \includegraphics[width=\linewidth,height=\textheight,keepaspectratio]{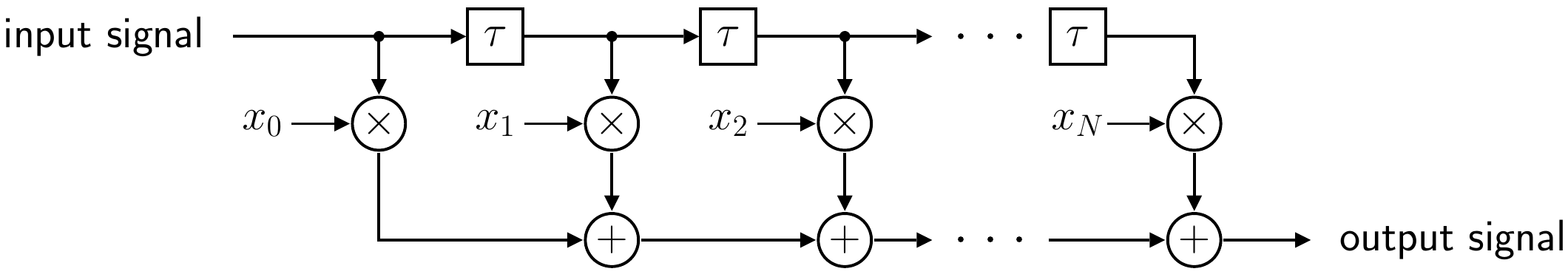}
    \caption{An example analog \sic filter (an $N$-tap \gls{fir} filter) with tunable tap weights $\braces{x_i}$ and fixed, uniform tap delay $\tau$.}
    \label{fig:asic}
\end{figure}

While it may seem fairly straightforward to implement analog \sic, it is practically quite challenging in most cases, especially outside of well-controlled lab settings.
Most notably, there is small margin for error in \sic due to the overwhelming strength of \si, reinforcing the need for extremely accurate, adaptable, and low-overhead \si reconstruction. 
Another challenging aspect is the need to miniaturize analog \sic filters into form-factors that integrate into devices such as cell phones, laptops, wireless routers, base stations, and the like \cite{kolodziej_techniques_2019}.
Miniaturization is especially challenging in settings where the delay spread of \si has the potential to be high, since propagation delays need to be physically realized within the analog \sic filter.







\subsection{Digital Self-Interference Cancellation}
Cancelling \si through digital signal processing is naturally an attractive option in addition to analog \sic.
The flexibility and sophistication of digital filtering can be applied to estimate and cancel \si and has had impressive success \cite{korpi_full-duplex_2014,korpi_widely_2014,korpi_dissertation}.
As depicted in \figref{fig:sic} and \figref{fig:asic-dsic}, digital \sic is executed after analog \sic and therefore aims to cancel residual \si that remains after prior efforts of cancellation.
Naturally, one may ask whether digital \sic can cancel all \si, rendering analog \sic unnecessary. 
In general, this is not possible for a few reasons, stemming from hardware limitations and nonidealities.

\runinhead{Limited Dynamic Range of Analog-to-Digital Converters.}
With reasonable resolution and appropriate gain control before analog-to-digital conversion, quantization noise is rarely an issue in traditional half-duplex systems.
In full-duplex systems, on the other hand, the strength of \si has the potential to \textit{saturate} \adcs, even with ideal \gls{agc} and a reasonable number of bits \cite{korpi_full-duplex_2014,ashu_inband_jsac_2014}. 
Since \gls{agc} acts on the combination of a desired signal plus \si and noise at the \adc input, the strength of quantization noise is dictated largely by that of \si.
In such cases, only a portion of the \adc's total dynamic range is effectively used to quantize a desired signal.
Consequently, even if \si could be completely reconstructed and cancelled digitally, its effects may remain in the form of increased quantization noise, which can severely and irreversibly degrade the quality of a desired receive signal.
This reinforces the need to sufficiently cancel \si before it reaches the \adc input, which is often most reliably done via analog \sic.


\runinhead{Transceiver Nonidealities.}
Practical transceivers introduce nonidealities in the transmit and receive chains, such as amplifier nonlinearities, I/Q imbalance, transmitter thermal noise, and phase noise, which complicate digital \sic since the digital domain does not have knowledge of these imperfections \cite{korpi_dissertation}.
When these nonidealities are not negligible, this requires digital \sic to accurately estimate and subsequently cancel them, which can be computationally complex.
To reduce this burden, analog \sic can be well positioned to cancel transmit-side impairments using the \rf transmit signal as input to its cancellation filter, which inherently will include nonidealities introduced by transmit \pas and transmit thermal noise, for example.
In addition, analog \sic can importantly ensure that the power of residual \si is sufficiently low such that it does not overwhelm and saturate receive-chain components such as \lnas, which practically have a limited dynamic range that can be exceeded by \si \cite{korpi_full-duplex_2014}.



\runinhead{Recent Breakthroughs using Machine Learning.}
In addition to classical signal processing approaches for digital \sic, solutions based on machine learning have been gaining traction and have shown impressive results \cite{stimming_implement_2018,guo_realtime_2018,stimming_nn_2018,stimming_implement_2020,guo_dsic_2019,mishra_mimo_2020,stimming_joint_2021}.
The main motivation for the use of machine learning over classical approaches for digital \sic is to capture transceiver nonidealities with reduced complexity.
Classical signal processing approaches have been able to effectively estimate and account for transceiver impairments when reconstructing and subsequently cancelling \si.
This is done by modeling transceiver impairments with established, parameterized models, but the estimation of model parameters is computationally expensive with classical methods. 
Machine learning has shown to be able to offer comparable performance as classical methods in capturing transceiver impairments when reconstructing \si but with reduced complexity.
Experimental validation of these digital \sic solutions based on machine learning has proven their effectiveness \cite{stimming_implement_2018,guo_realtime_2018,stimming_implement_2020}.
In addition, machine learning can be used to reduce the complexity of \sic in multi-antenna systems.
Rather than merely replicating single-antenna \sic solutions as an extension to multi-antenna systems, researchers have shown that machine learning can reduce the size and complexity of digital \sic by learning correlations between antennas \cite{mishra_mimo_2020}.




\subsection{Circulators and Antenna Isolation}
\begin{figure}[t]
    \centering
    \includegraphics[width=0.7\linewidth,height=\textheight,keepaspectratio]{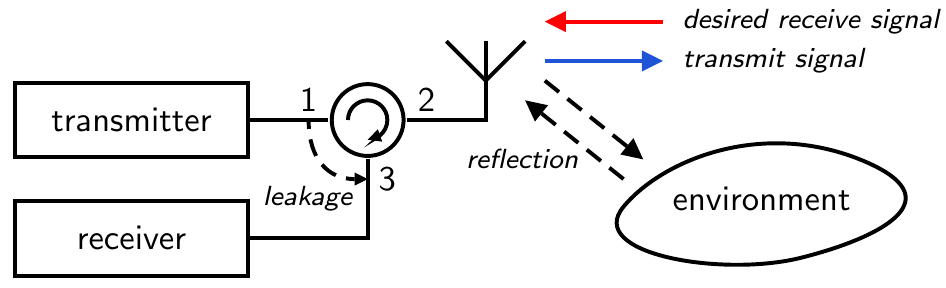}
    \caption{A circulator can be used to establish isolation between a transmitter and receiver sharing an antenna. Leakage through the circulator and reflections off the environment give rise to \si, however.}
    \label{fig:circulator}
\end{figure}

An \rf component known as a \textit{circulator} has been used in many monostatic radar and communication applications as a duplexer when a single antenna is used for simultaneous transmission and reception of RF signals \cite{kolodziej_techniques_2019,ashu_inband_jsac_2014}. 
In its simplest form, a circulator is a three-port, passive device where a signal entering a given port is ``circulated'' to the next port in the rotation. 
An example of this device being used with a single antenna shared by transmission and reception can be seen in \figref{fig:circulator}. 
Transmit signals enter Port 1 of the circulator and exit at Port 2, where they are radiated by the antenna.
Signals received by the antenna enter Port 2 and are circulated to Port 3, where they exit the circulator and enter the receive chain. 
This establishes isolation between the transmitter and receiver of a full-duplex device using a single antenna for transmission and reception.

One may reason that with an ideal circulator and with two radios operating in free space, full-duplex operation is trivial since perfect isolation is achieved between a radio's transmit signal and a desired receive signal. 
In reality, a circulator effectively offers limited \rf isolation between its ports, which introduces \si at the receiver. 
This is due to a number of factors, most notably the leakage between ports, reflections caused by imperfect matching at the antenna, and reflections off the environment.
Circulators with small form-factors that offer high isolation for full-duplex are an active area of research with immense potential \cite{laughlin_jsac_2014,dinc_2017}.
Nonetheless, analog and digital \sic can be used in conjunction with a circulator for single-antenna full-duplex transceivers.
In such cases, \sic aims to cancel circulator leakage, as well as reflections off the environment and from the antenna.


\begin{svgraybox}
    \textbf{Takeaways.}
    Digital \sic is an effective and flexible route to enabling full-duplex, but it is bottlenecked in practice by imperfections and limitations of hardware. 
    As such, it is often used in conjunction with analog \sic, which can inherently account for hardware impairments, relax the requirements of digital \sic, and prevent \si from saturating receive-chain components.
    Circulators and other duplexers can provide isolation between the transmitter and receiver of a full-duplex transceiver when using a single antenna, effectively weakening \si that analog and digital \sic must tackle.
\end{svgraybox}

\section{A New Frontier: Full-Duplex Millimeter-Wave Systems} \label{sec:mmwave}

To meet the ever-growing demand for high-rate wireless access, cellular networks have turned to \mmwave carrier frequencies, typically classified as ranging from around $30$ GHz to $100$ GHz \cite{jeff_5g_2014}.
Fifth-generation (5G) cellular networks and IEEE 802.11ad/ay, for instance, leverage frequency bands that span hundreds of megahertz.
These wide swaths of spectrum facilitate higher data rates and enable new applications in entertainment, industry, and sensing.
The widespread deployment of \mmwave networks has faced hurdles thus far, but it is expected they will see greater success through the end of the 2020s.
Both \mmwave communication systems and full-duplex technology were explored concurrently during the 2010s and were proposed as core technologies for next-generation wireless networks.
The combination of the two---full-duplex \mmwave communication systems---has not been explored as extensively.
Only recently has this topic garnered noteworthy attention from industry and academia \cite{roberts_wcm,xia_2017,liu_beamforming_2016,3GPP_IAB_2}.

\subsection{Prelude: Full-Duplex MIMO Transceivers} \label{subsec:mimo}

With multiple antennas at a transmitter and a receiver, there is the potential to multiplex more than one data stream over the resulting \mimo wireless channel via spatial signal processing \cite{heath_lozano}. 
\mimo communication transformed wireless networks forever by offering multiplicative rate gains over traditional \acrlong{siso} communication systems.
Given their prominence, the extension of full-duplex to \mimo transceivers was imperative.
With multiple antennas at the transmitter and receiver of a full-duplex transceiver, \si is inflicted onto each receive antenna by each transmit antenna, leading to a \mimo \si channel.
There is a quite natural extension of analog and digital \si to full-duplex \mimo transceivers \cite{korpi_dissertation,stanford_mimo_2014,Huberman_Le-Ngoc_2015}.
Perhaps more exciting, though, is the potential to mitigate \si through precoding and combining (i.e., spatial processing) at the transmitter and receiver of the full-duplex transceiver \cite{riihonen_loopback,himal_fdrelay_2014,ngo_jsac_2014,Day_Margetts_Bliss_Schniter_2012}. 
By strategically transmitting energy into the \si channel and receiving energy from it, \si can be mitigated spatially while still communicating desired signals in a \mimo fashion.
As the number of antennas grows---and especially in the massive \mimo regime---the prospects of spatial cancellation are even more promising.
There is extensive literature on the subject of full-duplex \mimo systems; we encourage interested readers to explore \cite{stanford_mimo_2014,Huberman_Le-Ngoc_2015,riihonen_loopback,himal_fdrelay_2014,ngo_jsac_2014,Day_Margetts_Bliss_Schniter_2012,everett_softnull_2016} and references therein for more details.
In the remainder of this section, we consider a particular class of full-duplex \mimo transceivers: those at \mmwave frequencies.
Solutions for full-duplex \mmwave systems draw inspiration from those for traditional full-duplex \mimo systems at sub-6 GHz but face unique challenges and are subject to new transceiver- and network-level factors.




\subsection{What's New at Millimeter-Wave Frequencies?}
Communication at \mmwave is more than a mere shift in carrier frequency, as elegantly stated in \cite{heath_overview_2016}.
In general, path loss increases with carrier frequency, which necessitates the use of dense antenna arrays to supply high beamforming gains that can deliver link margins that sustain high-rate communication.
Antenna arrays on \mmwave network infrastructure are typically equipped with $64$-$256$ antennas, whereas user equipment may be equipped with $4$-$16$ elements.
Fortunately, antenna footprint shrinks as carrier frequency increases, allowing dense antenna arrays to fit in convenient form-factors.
The severe path loss and susceptibility to blockage at \mmwave frequencies, coupled with highly directional beamforming, reduces inter-user interference and facilitates base station deployments much denser than traditional sub-6~GHz macrocell deployments.
All of this has led to new transceiver-level and network-level challenges and solutions at \mmwave.

\begin{figure}[t]
    \centering
    \includegraphics[width=0.64\linewidth,height=\textheight,keepaspectratio]{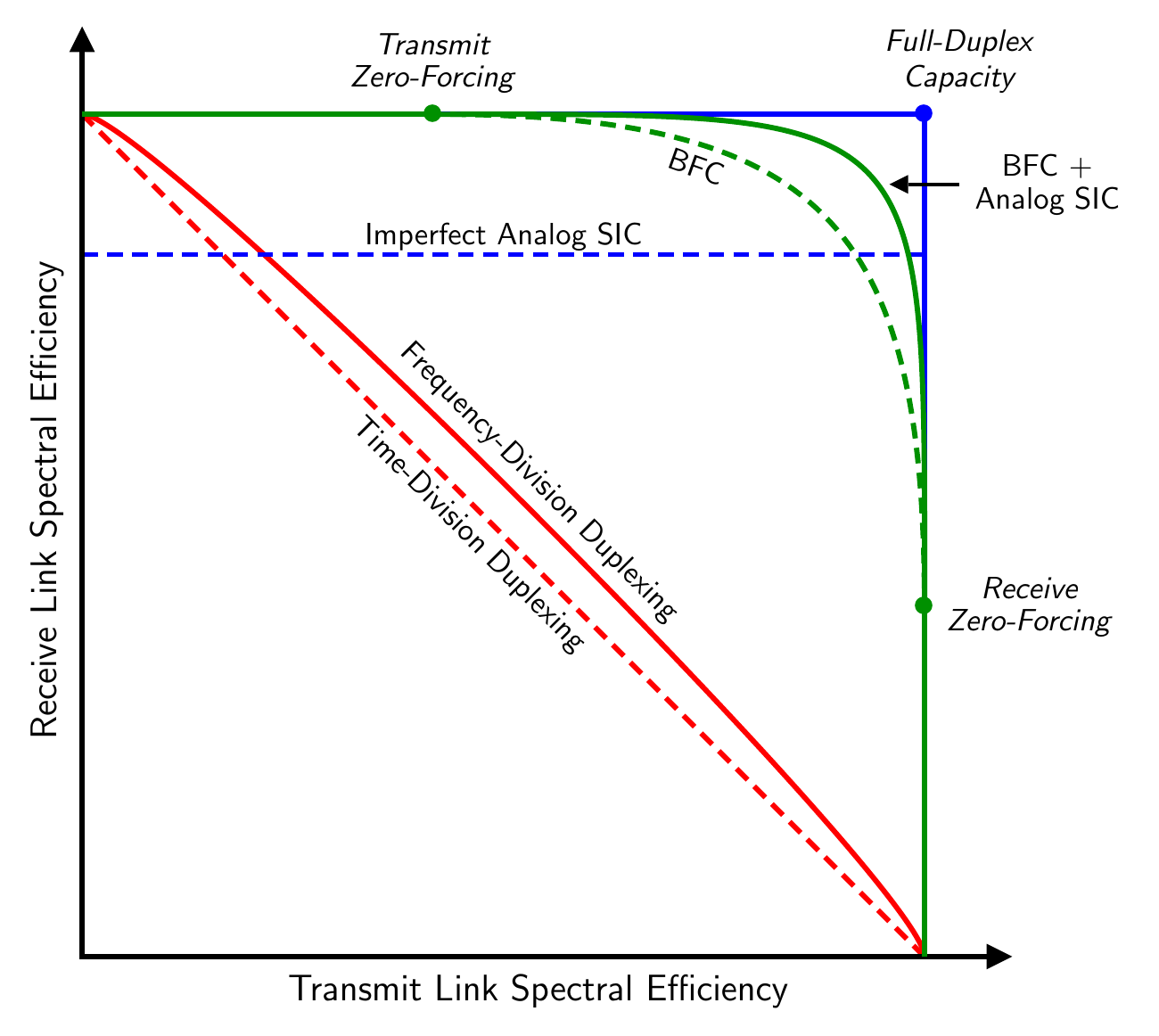}
    \caption{The rate region boundaries for \tdd, \fdd, and full-duplex using analog \sic, beamforming cancellation (shown as BFC), and the combination of the two. Beamforming cancellation consumes spatial resources, which introduces a gap between it and the full-duplex capacity. This gap can be reduced via analog and/or digital \sic. Reproduced from \cite{roberts_wcm} with permission.}
    \label{fig:regions}
\end{figure}

\subsection{Exciting Potential to Tackle Self-Interference via Beamforming} \label{subsec:bf}
In \subsecref{subsec:mimo}, we mentioned that a multi-antenna full-duplex system can design its transmit precoder and receive combiner to reduce \si coupled over the \mimo \si channel.
At \mmwave, the prospect of such \textit{spatial \sic} notably improves for a few reasons \cite{xia_2017,liu_beamforming_2016,roberts_wcm}.
First of all, path loss and blockage increases at \mmwave frequencies, compared to sub-6 GHz frequencies, which presumably weakens \si as it couples from a transmitter to receiver, both directly over-the-air and due to reflections off the environment; reflectivity increases at \mmwave, however, which may lead to more \si from the environment.
Second, with denser antenna arrays, the degrees-of-freedom available to cancel \si increases, compared to traditional sub-6 GHz \mimo systems, which typically only have $2$-$8$ antennas.
It is important to note that, as the number of antennas has increased immensely at \mmwave, the number of data streams communicated has remained of similar order.
Hybrid analog/digital beamforming and analog-only beamforming architectures, which are ubiquitous thus far in \mmwave communication systems, fundamentally limit the number of data streams to the number of \dacs and \adcs, which is comparable to that in conventional sub-6 GHz \mimo systems.
With one or few data streams communicated and tens or hundreds of antennas, many degrees-of-freedom are potentially available for cancelling \si via strategic design of transmit and receive beamformers at a full-duplex \mmwave transceiver.

Excitedly, if beamforming can cancel \si sufficiently, the need for analog and/or digital \sic vanishes, meaning there is the potential for full-duplex \mmwave systems to operate without any additional hardware or complex signal processing.
In other words, analog and digital \sic may be needed for full-duplex \mmwave systems with sufficient spatial \sic.
Plenty of existing work has highlighted this by designing transmit and receive beamformers in such a way that \si is mitigated while maintaining downlink and uplink to users \cite{roberts_wcm,riihonen_loopback,everett_softnull_2016,xia_2017,liu_beamforming_2016,roberts_2021_robustcb,lopez-valcarce_beamformer_2019,lopez_prelcic_2019_analog,prelcic_2019_hybrid,satyanarayana_hybrid_2019,zhu_uav_joint_2020,da_silva_2020,lopez_analog_2022,koc_ojcoms_2021,cai_robust_2019,roberts_bflrdr,roberts_equipping_2020}.
Interestingly, unlike traditional analog and digital \sic solutions, transmit beamforming introduces the unique opportunity to reduce the degree of \si that ever reaches the receive antennas.
Like analog \sic, transmit and receive beamforming can be used to prevent \si from saturating receive chain components. 
For instance, in \cite{roberts_bflrdr}, a hybrid beamforming design is presented that guarantees \si is below some power level at each \lna and each \adc at the receiver of the full-duplex \mmwave device, ensuring they do not saturate. 
In addition to using beamforming alone to mitigate \si, researchers have also considered analog and/or digital \sic in conjunction, which relaxes the cancellation requirements of beamforming, allowing it to better serve uplink and downlink at the cost of added hardware or signal processing \cite{Huberman_Le-Ngoc_2015,roberts_equipping_2020,smida_asic,Alexandropoulos_Duarte_2017,singh_2020_acm,Zhang_Luo_2019,dinc_60_2016,suk_iab_2022}.
The rate region boundaries with spatial cancellation via beamforming (sometimes termed \textit{beamforming cancellation}) and analog \sic are shown in \figref{fig:regions} \cite{roberts_wcm}.

\subsection{Example Beamforming Design Problems}
We now overview two sample spatial \sic design problems for full-duplex \mmwave systems, with the goal of introducing readers to the design objectives and considerations surrounding such research problems and those taking a similar form.

\begin{figure}[t]
    \centering
    \includegraphics[width=0.65\linewidth,height=\textheight,keepaspectratio]{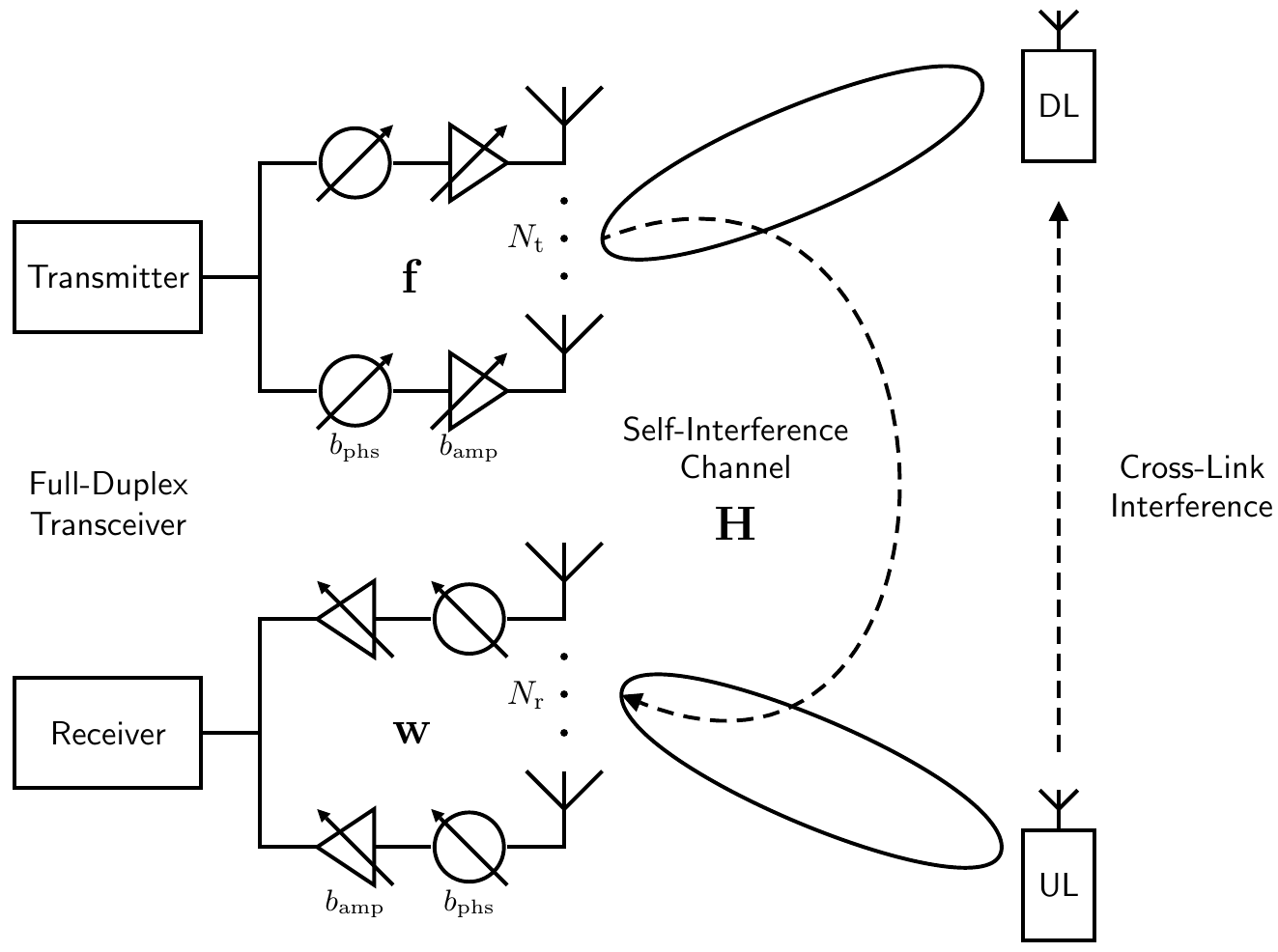}
    \caption{A full-duplex \mmwave transceiver transmits to a downlink user while receiving from an uplink user in-band. By strategically constructing its analog beamformers $\vf$ and $\vw$, it can reduce the level of \si coupled across the \mimo channel $\mH$ while delivering service to the users.}
    \label{fig:mmwave}
\end{figure}

\runinhead{Analog Beamforming Design.} 
Consider the system illustrated in \figref{fig:mmwave}.
Suppose the full-duplex \mmwave \bs equipped with analog-only beamforming---as opposed to hybrid digital/analog beamforming \cite{heath_overview_2016}---is serving an uplink user and a downlink user simultaneously and in-band, both of which are single-antenna devices.
It is currently practical to consider the use of separate, independently-controlled transmit and receive arrays at the \bs equipped with $\Nt$ and $\Nr$ antennas, respectively \cite{roberts_wcm}.
Let $\vf \in \setmatrixcomplex{\Nt}{1}$ and $\vw \in \setmatrixcomplex{\Nr}{1}$ be the transmit and receive beamforming vectors used at the full-duplex \bs. 
Let $\mH \in \setmatrixcomplex{\Nr}{\Nt}$ be the \mimo \si channel matrix manifesting between the transmit and receive arrays.
Let $\vhtx\ctrans \in \setmatrixcomplex{1}{\Nt}$ be the channel vector from the \bs to the downlink user.
Let $\vhrx \in \setmatrixcomplex{\Nr}{1}$ be the channel vector from the uplink user to the \bs.
The full-duplex \bs will rely solely on beamforming to mitigate \si with no additional analog or digital \sic.

Practical analog beamforming networks are comprised of digitally-controlled phase shifters.
In addition to phase control, some networks also offer quantized amplitude control through digitally-controlled attenuators or \glspl{vga} (see \subsecref{subsec:challenges}).
Quantized phase and amplitude control confines all physically realizable analog beamformers to come from some discrete sets, which can be captured as simply $\vf \in \set{F}$ and $\vw \in \set{W}$.
The transmit link and receive link \gpsnr are functions of their beamformers and can be expressed as
\begin{align}
\snrtx\parens{\vf} = \frac{\powertxbs \cdot \bars{\vhtx\ctrans \vf}^2}{\powernoiseue}, \qquad
\snrrx\parens{\vw} = \frac{\powertxue \cdot \bars{\vw\ctrans \vhrx}^2}{\powernoisebs},
\end{align}
where $\powertxbs$ and $\powertxue$ are the transmit powers of the \bs and the uplink user, while $\powernoisebs$ and $\powernoiseue$ are the noise powers of the \bs and the downlink user.
The \si and cross-link interference terms of the system are
\begin{align}
\inrrx\parens{\vf,\vw} = \frac{\powertxbs \cdot \bars{\vw\ctrans \mH \vf}^2}{\powernoisebs}, \qquad
\inrtx = \frac{\powertxue \cdot \bars{\hcl}^2}{\powernoiseue},
\end{align}
where $\inrrx$ is a function of the transmit and receive beams at the full-duplex \bs and $\inrtx$ is solely a function of the cross-link channel $\hcl$ from the uplink user to the downlink user.
Together, these desired and interference terms form the \gpsinr of the two links as
\begin{align}
\sinrtx\parens{\vf} = \frac{\snrtx\parens{\vf}}{1 + \inrtx}, \qquad
\sinrrx\parens{\vf,\vw} = \frac{\snrrx\parens{\vw}}{1 + \inrrx\parens{\vf,\vw}},
\end{align}
which determine the achievable spectral efficiencies as
\begin{align}
\setx\parens{\vf} = \logtwo{1 + \sinrtx\parens{\vf}}, \qquad
\serx\parens{\vf,\vw} = \logtwo{1 + \sinrrx\parens{\vf,\vw}},
\end{align}
when treating interference as noise.
Notice that the transmit link quality $\sinrtx$ and therefore its spectral efficiency $\setx$ are solely functions of the transmit beam $\vf$.
The fate of the receive link, however, is determined by both $\vw$ and $\vf$ due to \si.
It would desirable to design $\vf$ and $\vw$ such that they deliver high \gpsnr and couple low \si.

This motivates the following analog beamforming design problem, which aims to maximize the sum spectral efficiency of the system, while requiring the transmit and receive beams to be physically realizable.
\begin{subequations}
\begin{align} \label{eq:problem-mmwave-bf}
\max_{\vf,\vw} \ & \setx\parens{\vf} + \serx\parens{\vf,\vw} \\
\st 
& \vf \in \set{F}, \vw \in \set{W}
\end{align}
\end{subequations}
Several existing works on \mmwave full-duplex aim to solve this problem or one of similar form \cite{lopez_analog_2021,lopez_analog_2022,da_silva_2020}.
In general, this problem is difficult to solve due to the non-convexity of the objective from the coupling of $\vf$ and $\vw$ and the fact that $\set{F}$ and $\set{W}$ are non-convex sets.
Researchers typically instead tackle problems that are more readily solved and still yield high sum spectral efficiency.

One practical issue with solving analog beamforming problems of this type is that they are executed for each user pair, which can consume prohibitive amounts of radio resources (e.g., for channel estimation and over-the-air feedback) and computational resources.
In fact, the time required to solve these sorts of problems may not translate to timescales of practical systems, even with modern computing power.
Moreover, many existing solutions rely on unrealistic assumptions, such as real-time downlink/uplink \mimo channel knowledge (i.e., $\vhtx\ctrans$ and $\vhrx$), which is not obtainable in practical \mmwave systems today.



\runinhead{Analog Beamforming Codebook Design.}
In the previous example, we considered the problem of designing transmit and receive beams that maximize sum spectral efficiency. 
Now, to circumvent and overcome some of the practical challenges mentioned in the previous example, let us consider the goal of designing transmit and receive \textit{codebooks} that maximize sum spectral efficiency in full-duplex \mmwave systems.
Let us build on the previous example and the notation used therein.
Suppose it is required that the transmit beam $\vf$ be selected from some set of $\Mtx$ transmit beams $\braces{\vf_1,\vf_2,\dots,\vf_{\Mtx}}$, called a transmit codebook, which is common in practical systems. 
Likewise, suppose it is required that the receive beam $\vw$ be selected from some codebook of $\Mrx$ receive beams $\braces{\vw_1,\vw_2,\dots,\vw_{\Mrx}}$.
Here, $\Mtx$ and $\Mrx$ are on the order of tens or hundreds at most, meaning it is realistically the case that $\Mtx \ll \card{\set{F}}$ and $\Mrx \ll \card{\set{W}}$. 

\begin{figure}[t]
    \centering
    \includegraphics[width=\linewidth,height=0.2\textheight,keepaspectratio]{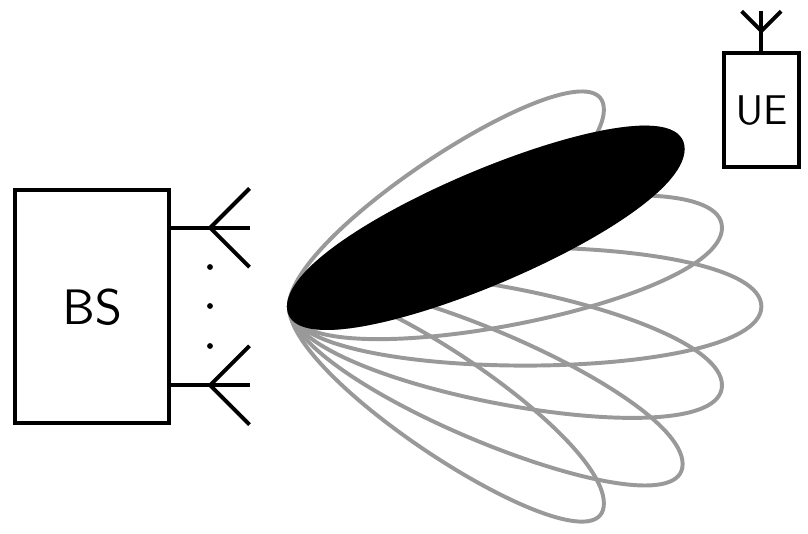}
    \caption{A \mmwave \bs conducts beam alignment by sweeping a codebook of candidate beams, selecting that which maximizes \gsnr to serve a given user.}
    \label{fig:alignment}
\end{figure}

It is fairly straightforward to design transmit and receive codebooks for traditional half-duplex \mmwave systems, typically done by tessellating beams to cover a desired service region to ensure that a user falling in this region can be served with high gain with at least one beam from the codebook (e.g., see \figref{fig:alignment}).
The design of codebooks for full-duplex \mmwave systems, on the other hand, is far more involved and has only been investigated in \cite{roberts_lonestar,roberts_2021_robustcb} thus far. 
Consider the following design problem, which aims to design codebook matrices $\mF \in \setmatrixcomplex{\Nt}{\Mtx}$ and $\mW \in \setmatrixcomplex{\Nr}{\Mrx}$ that maximize the expected sum spectral efficiency across a known user distribution for a given $\mH$, for instance.
\begin{subequations}%
\begin{align}
\max_{\mF,\mW} \ & \ \expect\brackets{\max_{\vf,\vw} \ \setx\parens{\vf} + \serx\parens{\vf,\vw}} \\
\st  
& \ \vf = \entry{\mF}{:,i}, \vw = \entry{\mW}{:,j} \\
& \ \entry{\mF}{:,i} \in \precb \ \forall \ i = 1, \dots, \Mtx \label{eq:cb-phys-F} \\
& \ \entry{\mW}{:,j} \in \comcb \ \forall \ j = 1, \dots, \Mrx \label{eq:cb-phys-W}
\end{align}
\end{subequations}
Here, these codebook matrices are structured such that the $i$-th column of $\mF$ is $\vf_i$ and the $j$-th column of $\mW$ is $\vw_j$, both of which are required to be physically realizable beamforming vectors, hence \eqref{eq:cb-phys-F} and \eqref{eq:cb-phys-W}.
To serve each user pair, some transmit and receive beams $\vf$ and $\vw$ are selected from their respective codebooks. 
With random user placement, any transmit and receive beam may be chosen from the codebooks, meaning desirable codebooks $\mF$ and $\mW$ would offer low $\inrrx\parens{\vf,\vw}$ for all possible $\vf$ and $\vw$ while still capable of delivering high beamforming gains.
This is a difficult problem to solve, largely due to the daunting objective of aiming to maximize average sum spectral efficiency, but provides good direction for desirable codebooks $\mF$ and $\mW$.

\begin{figure}[t]
    \centering
    \includegraphics[width=0.47\linewidth,height=\textheight,keepaspectratio]{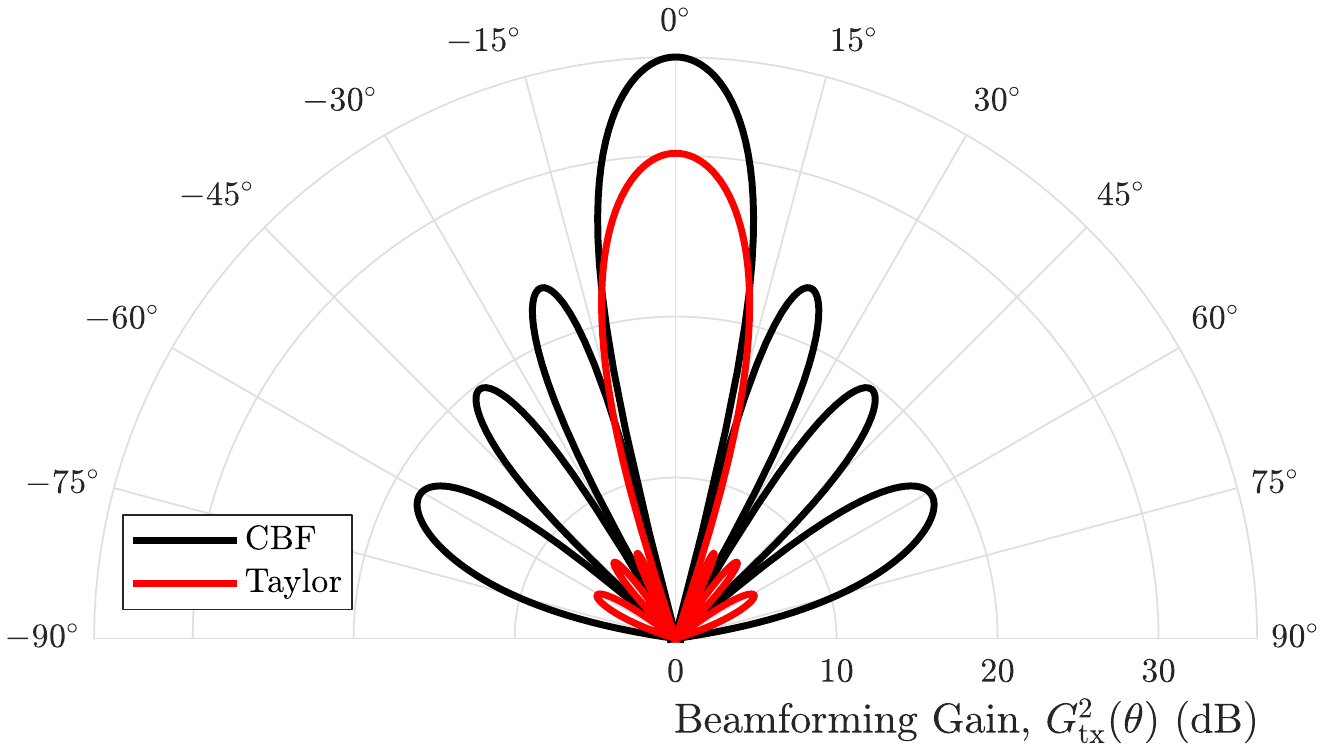}
    \qquad
    \includegraphics[width=0.47\linewidth,height=\textheight,keepaspectratio]{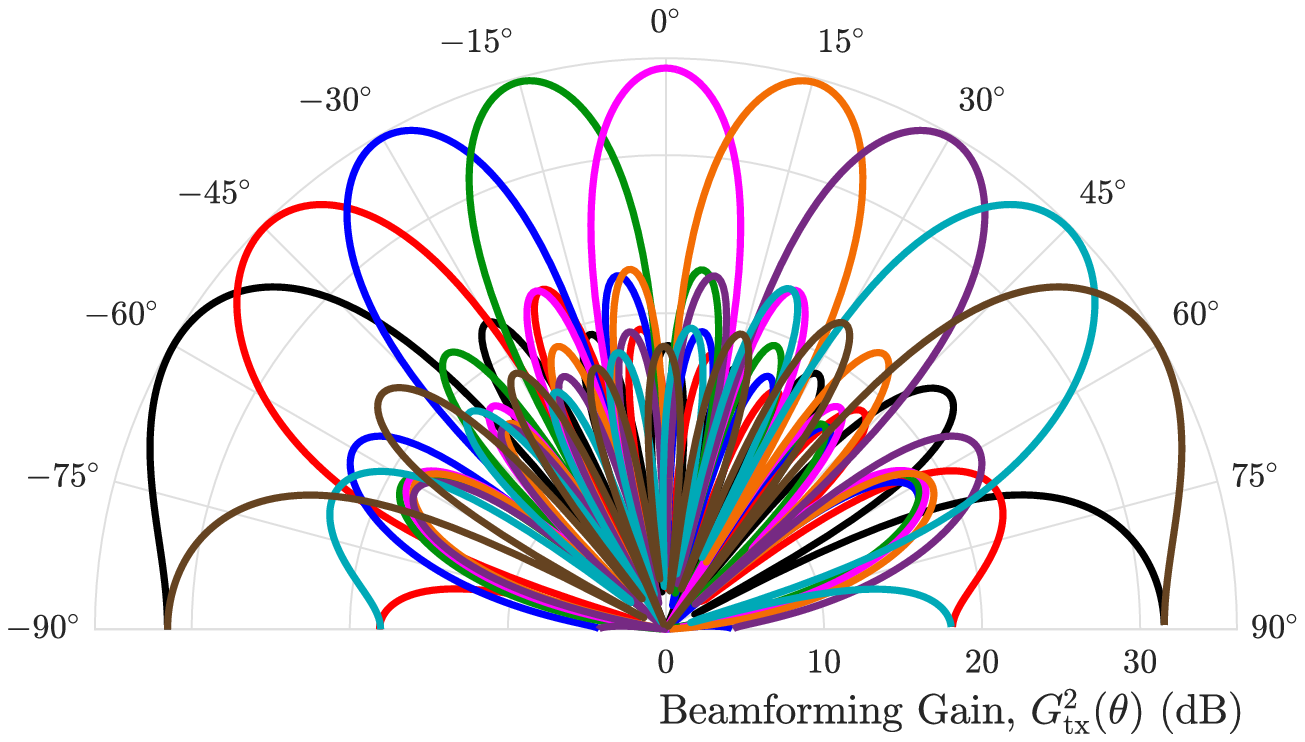}   
    \caption{(left) A broadside beam from two conventional beamforming codebooks. (right) Beams created by \cite{roberts_lonestar,roberts_2021_robustcb} that span a coverage region with high gain while coupling low \si.}
    \label{fig:codebooks}
\end{figure}

In \cite{roberts_lonestar,roberts_2021_robustcb}, a similar problem is tackled, as shown below in problem \eqref{eq:lonestar}.
Here, the objective is instead to minimize average \si coupled between possible transmit and receive beams across the channel $\mH$, effectively minimizing average $\inrrx\parens{\vf,\vw}$.
In doing so, high beamforming gain and broad coverage over some transmit and receive coverage regions are enforced by \eqref{eq:cb-gain-F} and \eqref{eq:cb-gain-W}, where $\sigmatxsq$ and $\sigmarxsq$ are design parameters that throttle the so-called coverage variance of each codebook.
In essence, these constraints ensure codebooks can reliably deliver high $\snrtx$ and high $\snrrx$, while the objective aims to minimize $\inrrx$.%
\begin{subequations}\label{eq:lonestar}
\begin{align}
\min_{\mF,\mW} \ & \ \normfro{\mW\ctrans \mH \mF}^2 \\
\st  
& \ \normtwo{\Nt\cdot\vone - \diag{\Atx\ctrans\mF}}^2 \leq \sigmatxsq \cdot \Nt^2 \cdot \Mtx \label{eq:cb-gain-F} \\
& \ \normtwo{\Nr\cdot\vone - \diag{\Arx\ctrans\mW}}^2 \leq \sigmarxsq \cdot \Nr^2 \cdot \Mrx \label{eq:cb-gain-W} \\
& \ \entry{\mF}{:,i} \in \precb \ \forall \ i = 1, \dots, \Mtx \\
& \ \entry{\mW}{:,j} \in \comcb \ \forall \ j = 1, \dots, \Mrx
\end{align}
\end{subequations}%
A codebook of beams output by this design is shown in \figref{fig:codebooks}.
Compared to traditional beams, the beams produced by \cite{roberts_lonestar,roberts_2021_robustcb} make use of side lobes to cancel \si spatially while providing adequate coverage across the service region from $-60^\circ$ to $60^\circ$.
The codebooks designed with this framework proved to offer far greater robustness to \si and similar beamforming gain when compared to conventional codebooks, which allowed them to deliver sum spectral efficiencies $\setx + \serx$ that approach the full-duplex capacity without analog or digital \sic.
Designs like this are particularly exciting because they have the potential to accommodate codebook-based beam alignment while also mitigating \si through beamforming.
For more details and more extensive evaluation of this design, please see \cite{roberts_lonestar,roberts_2021_robustcb}.

\subsection{Key Practical Challenges and Considerations} \label{subsec:challenges}
We now outline important considerations when designing solutions for practical full-duplex \mmwave systems, some of which have already been touched on in this chapter.

\runinhead{Digitally-Controlled Analog Beamforming Networks.}
Unlike digital beamforming, which takes place in software/logic, analog beamforming is executed using phase shifter components, potentially along with attenuators and/or amplifiers, all of which are digitally-controlled.
In other words, some finite number of bits are dedicated to realizing the phase and amplitude of each beamforming weight.
For instance, the discrete set of physically realizable phase settings by phase shifters with settings uniformly distributed between $0$ and $2\pi$ with resolution $\bitsphase$ bits can be expressed as
\begin{align}
\set{P} &= \braces{\theta_i = \frac{(i-1) \cdot 2 \pi}{2^\bitsphase} : i = 1, \dots, 2^\bitsphase},
\end{align}
meaning it is practically required, for the $i$-th beamforming weight $w_i$, that $\angleop{w_i} \in \set{P}$ for all $i$.
In practice, it should be noted that both phase and amplitude control typically have some error associated with them, which is generally frequency-dependent.

Some phased arrays employ phase shifters based on a vector modulator architecture, where the in-phase and quadrature components of a signal can be scaled independently to produce a desired phase shift, in which case phase shifter settings are presumably no longer uniformly spaced.
Note that such an architecture also offers a means to scale the amplitude of the output signal. 
Practical beamforming-based full-duplex solutions for \mmwave systems should account for the limitations imposed by a particular analog beamforming architecture.
Quantized control of each beamforming weight leads to non-convex sets that are difficult to optimize over. 
Blindly projecting a solution onto the set of physically realizable beamforming vectors can be detrimental, as small errors in this full-duplex setting are magnified by the sheer strength of \si relative to a desired signal.
This motivates the use of high-resolution phase shifters in full-duplex \mmwave systems and/or ways to handle the non-convexity posed by quantized phase shifters.




\runinhead{Accommodating Codebook-Based Beam Alignment.}
Codebook-based analog beamforming is a critical component of \mmwave communication systems \cite{heath_overview_2016,ethan_beam}.
Rather than measure a high-dimensional \mimo channel and subsequently configure an analog beamforming network, modern \mmwave systems instead rely on beam alignment procedures to identify promising beamforming directions, typically via exploration of a codebook of candidate beams \cite{heath_overview_2016,ethan_beam,junyi_wang_beam_2009}.
This offers a simple and robust way to configure an analog beamforming network without downlink/uplink \mimo channel knowledge \textit{a priori}, which is not obtainable in practice. 

Let $\precbbar$ and $\comcbbar$ be transmit and receive beamforming codebooks used at a full-duplex device.
Executing beam alignment on each link independently (in a half-duplex fashion) would aim to solve (or approximately solve) the following problems or ones taking a similar form.
\begin{align} \label{eq:beam-alignment}
\vf\opt = \argmax_{\vf \in \precbbar} \ \snrtx\parens{\vf}, \qquad
\vw\opt = \argmax_{\vw \in \comcbbar} \ \snrrx\parens{\vw}
\end{align}
If selecting the transmit and receive beams independently to maximize each link's \gsnr, the selected beam pair may couple high \si when using traditional beamforming codebooks.
In other words, $\inrrx\parens{\vf\opt,\vw\opt}$ may be much greater than $0$~dB.
In fact, we show this has been confirmed by recent measurements \cite{roberts_att_angular}, which we cover shortly in \subsecref{subsec:experimental}.
Therefore, one can imagine it would be preferable from a full-duplex perspective to jointly select transmit and receive beams that deliver high $\snrtx$ and $\snrrx$ and also couple low $\inrrx$.
This is precisely the motivation for \cite{roberts_steer}, which is also introduced in \subsecref{subsec:experimental}.
If codebooks could be designed such that all possible $\parens{\vf\opt,\vw\opt}$ couple sufficiently low \si, then beam alignment may be conducted on the transmit and receive links independently, as shown in \eqref{eq:beam-alignment}, with guarantees of low \si regardless of which beams are selected---the motivation for \cite{roberts_lonestar,roberts_2021_robustcb} and the codebook design problem introduced in \subsecref{subsec:bf}.
Creating solutions like these that accommodate beam alignment will be critical to the deployment of full-duplex \mmwave systems.


\runinhead{Self-Interference Channel Estimation and Limited Channel Knowledge.}
As mentioned just previously, current practical \mmwave systems circumvent downlink/uplink channel estimation via beam alignment, meaning they do not have knowledge of the transmit and receive \mimo channels $\vhtx$ and $\vhrx$.
Practical beamforming-based solutions for full-duplex \mmwave systems should account for this. 
Efficient and accurate estimation of the \si \mimo channel $\mH$ is a research problem still in its infancy.
This is perhaps most largely due to the fact that modeling $\mH$ is still an open research problem itself, whose outcomes may inspire strategies for its estimation.
\mimo channel estimation in \mmwave transceivers is complicated by the sheer size of these channels and the fact that \dacs and \adcs observe the channel through the lens of analog beamformers \cite{heath_overview_2016}.
Routes to reduce estimation overhead would be valuable contributions, potentially by leveraging static portions of the \si channel (e.g., the direct coupling between the arrays) and/or by accurate channel modeling.
Nonetheless, whatever \si channel estimation strategies are developed will naturally be imperfect to some degree, suggesting that practical designs should be robust to channel estimation error.
Robustness is especially critical in full-duplex settings, since small errors in mitigating \si can be detrimental due to its overwhelming strength.



\runinhead{Leveraging User Selection.}
A full-duplex \mmwave \bs will likely serve multiple downlink users and uplink users over many time slots.
As assumed thus far, let us consider the case where the \bs can serve a single downlink-uplink user pair in a full-duplex fashion at any given time, multiplexing user pairs in time. 
The degree of \si coupled at the full-duplex \bs depends on the transmit and receive beams when serving a particular downlink-uplink user pair. 
In addition, the degree of cross-link interference depends on the two users being served.
When given a pool of user pairs needing service, one can therefore imagine that strategically selecting which user pair to serve has the potential to be a powerful tool to improve full-duplex performance \cite{roberts_wcm}.
This concept has not been fully fleshed out in existing literature and deserves future study.
Interesting future work, for instance, would be the design of intelligent schedulers that incorporate full-duplex user selection to improve link-level spectral efficiency and gains in network-level throughput.

\begin{figure}[t]
    \centering
    \includegraphics[width=0.7\linewidth,height=\textheight,keepaspectratio]{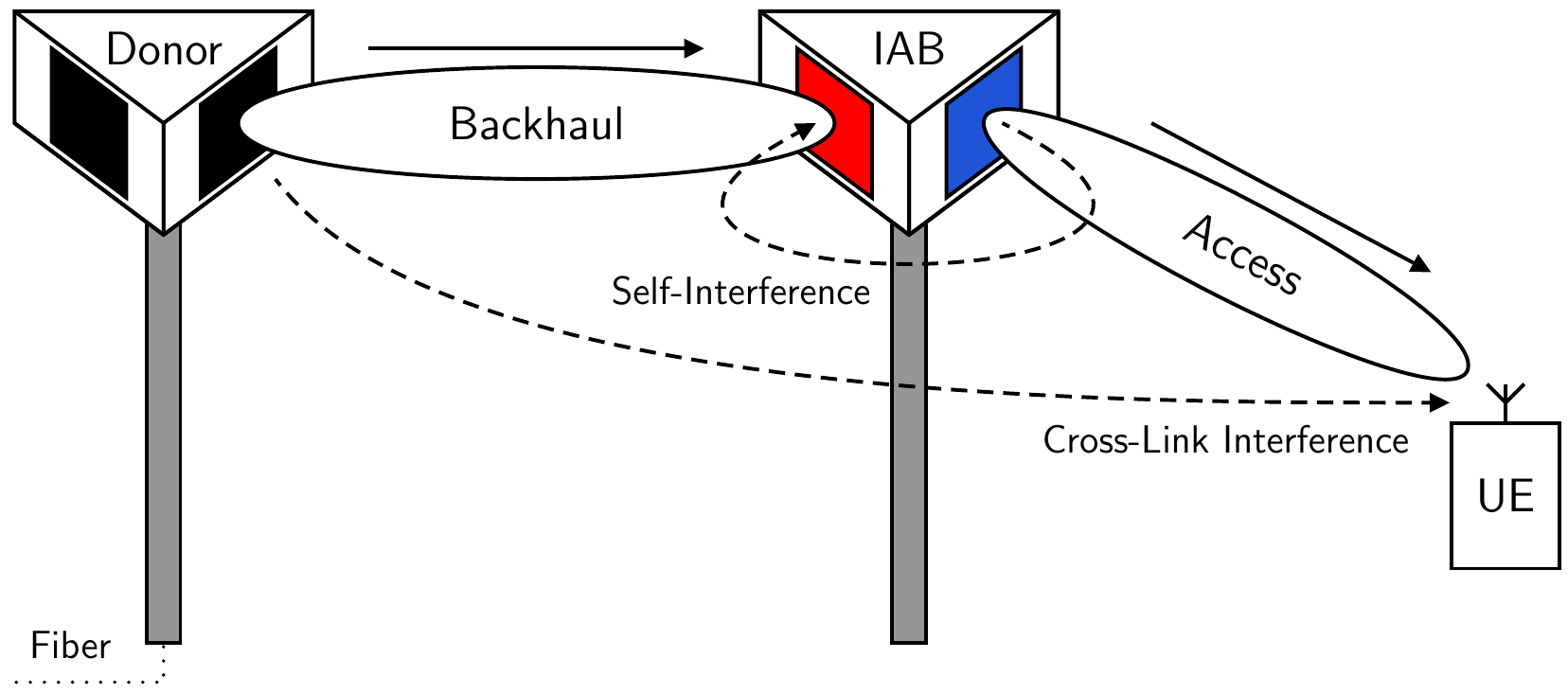}
    \caption{A full-duplex \mmwave \iab node receives wireless backhaul from a fiber-connected donor on one sector while transmitting access to a downlink user on another sector.}
    \label{fig:iab}
\end{figure}

\subsection{Full-Duplex Integrated Access and Backhaul} \label{subsec:iab}
A particularly motivating application of full-duplex in \mmwave cellular networks is in \acrfull{iab} \cite{cudak21integrated,iab,3GPP_IAB_2}, where a fiber-connected \bs serves nearby users and is responsible for maintaining a wireless backhaul to one or more nearby \bss, as illustrated in \figref{fig:iab}.
By using the same pool of \mmwave spectrum for access and wireless backhauling, dense \mmwave networks can be deployed with fewer dedicated fiber connections, which reduces the cost, time, and permitting associated with deployment.
Like other multi-hop wireless networks, however, \iab networks have faced scaling challenges due to degraded throughput and higher latency as the network grows.
Recent work has investigated the use of full-duplex to overcome these obstacles \cite{gupta_fdiab,suk_iab_2022}.

\begin{figure}[t]
    \centering
    \includegraphics[width=\linewidth,height=0.13\textheight,keepaspectratio]{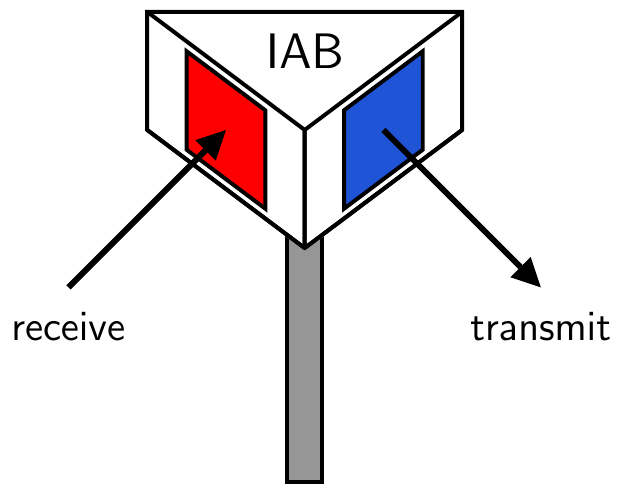}
    \qquad\qquad
    \includegraphics[width=\linewidth,height=0.13\textheight,keepaspectratio]{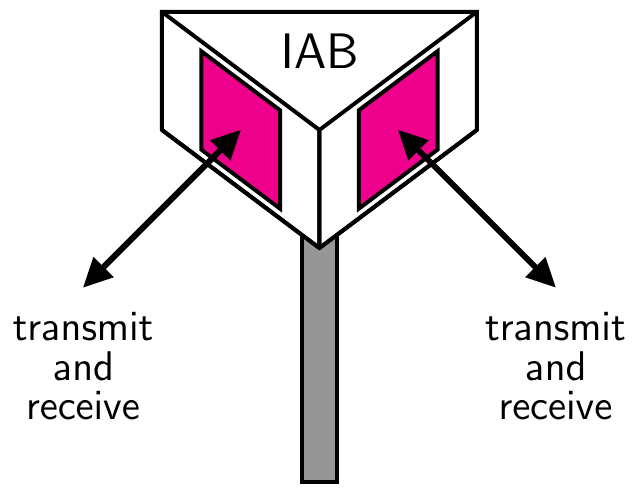}
    \caption{(left) An inter-sector full-duplex \iab node full-duplexes transmission and reception across sectors. (right) An intra-sector full-duplex \iab node full-duplexes transmission and reception within each sector.}
    \label{fig:inter-intra}
\end{figure}






\runinhead{Inter-sector and intra-sector full-duplex.} 
An \iab node can operate in a full-duplex fashion in two main ways, depending on scope. 
Consider a sectorized \iab node with three sectors, each equipped with a transceiver serving a $120^\circ$ field-of-view.
The first potential full-duplex operating mode, which we refer to as \textit{inter-sector} full-duplex, allows each sector to either transmit or receive, meaning \si may be inflicted by one sector onto one or both of the other sectors.
\figref{fig:iab} depicts inter-sector full-duplex, for instance, where one sector receives while another transmits in-band.
With inter-sector full-duplex, sectors are no longer required to collectively transmit or collectively receive but rather can be scheduled independently, and the transceiver on each sector can be merely half-duplex-capable. 
This full-duplex mode unlocks scheduling opportunities that are otherwise not available, allowing the network to achieve higher throughput, as we will highlight shortly.

The other potential full-duplex operating mode we refer to as \textit{intra-sector} full-duplex, where each sector is equipped with a full-duplex transceiver, and we illustrate in \figref{fig:inter-intra}.
In this case, each sector may transmit and receive from a downlink-uplink user pair within its field-of-view.
Notice that, when each sector operates simultaneously and in-band as its neighboring sectors, \si is inflicted from each transmitting sector onto each receiving sector.
Naturally, intra-sector full-duplex has the potential to outperform inter-sector full-duplex, but the gains of such have not been thoroughly investigated.
Early deployments of full-duplex \mmwave \bss, especially for \iab, will likely be of the inter-sector form, since \si is likely less severe and half-duplex transceivers can be used per sector.

\runinhead{Recent Progress Validating Full-Duplex IAB.}
Full-duplex \iab networks are studied in \cite{gupta_fdiab} to characterize the throughput and latency gains when upgrading \iab nodes from half-duplex to full-duplex transceivers.
Note that, in \cite{gupta_fdiab}, users were kept as half-duplex devices, which is a realistic assumption for the foreseeable future.
The authors show through analysis and simulation that, with full-duplex \iab nodes, user latency can reduce four-fold and user throughput can improve eight-fold for fourth-hop users---far transcending the familiar doubling of spectral efficiency offered by full-duplex at the link level.
In general, \cite{gupta_fdiab} shows that users further from the donor enjoy greater performance improvements with full-duplex.
This can be explained by the fact that full-duplex \iab networks can meet latency and throughput targets that half-duplex \iab networks cannot, yielding relative gains that can tend to infinity.
Ultimately, the gains offered by full-duplex are thanks to the scheduling opportunities it unlocks: certain links that must be orthogonalized with half-duplex \iab nodes need not be with full-duplex, allowing packets to more quickly propagate through the multi-hop network.
Compared to their half-duplex counterparts, full-duplex \iab networks can facilitate reduced latency, higher throughput, fairer service, and deeper networks---even with imperfect \sic \cite{gupta_fdiab}.




\begin{figure}[t]
    \sidecaption[t]
    \centering
    \includegraphics[width=0.55\linewidth,height=\textheight,keepaspectratio]{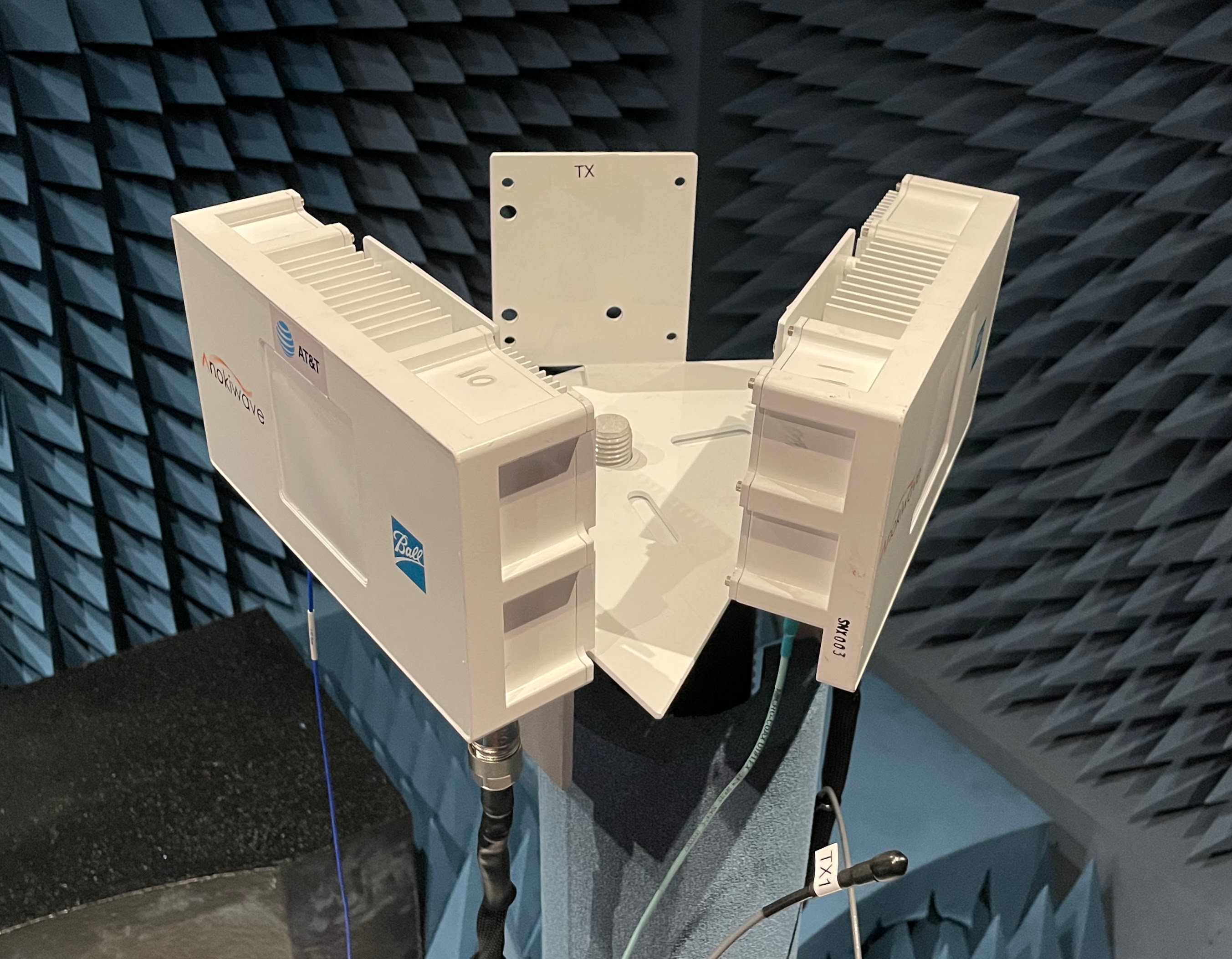}
    \caption{The 28 GHz phased array platform used for measurements of \si in \cite{roberts_att_angular,roberts_2021_att_measure,roberts_steer}. Transmit array on right; receive array on left. Received \si power depends on the steering direction of the transmit and receive beams. The multi-panel triangular platform shown here is a relevant deployment option for \mmwave small-cell \bss and \iab nodes. Reproduced from \cite{roberts_att_angular} with permission.}
    \label{fig:setup}
\end{figure}

\subsection{Recent Experimental Research Outcomes} \label{subsec:experimental}
The majority of research on full-duplex \mmwave systems has been theoretical in nature, using simulation to validate proposed solutions.
Recently, there has been work experimentally investigating full-duplex \mmwave systems, two efforts of which we introduce herein.

\begin{figure}[t]
    \sidecaption[t]
    \centering
    \includegraphics[width=0.6\linewidth,height=\textheight,keepaspectratio]{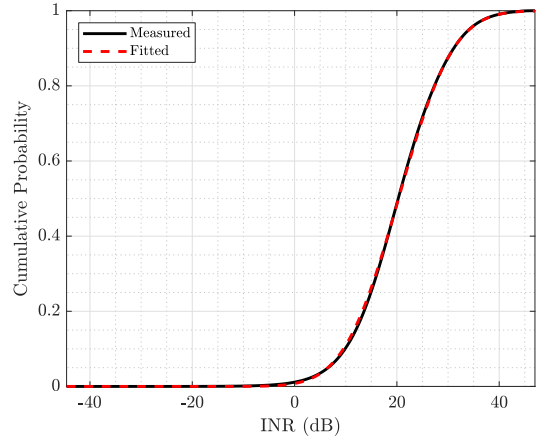}
    \caption{The empirical \gls{cdf} of the nearly 6.5 million measurements of 28 GHz \si using 16$\times$16 planar arrays in \cite{roberts_att_angular}, along with a fitted log-normal distribution. Less than 1\% of transmit-receive beam pairs yield $\inrrx \leq 0$ dB. Highly directional \mmwave beams do not necessarily offer sufficient isolation for full-duplex but strategically selecting them can. Reproduced from \cite{roberts_att_angular} with permission.}
    \label{fig:cdf-inr}
\end{figure}

\runinhead{Measurements of mmWave Self-Interference.}
\si was studied quite extensively over the past decade or so, largely in the context of sub-6 GHz transceivers.
To progress development of full-duplex \mmwave systems, a necessary first step is to better understand \si in \mmwave systems.
Measurements of \mmwave \si in \cite{suk_iab_2022,rajagopal_2014,kohda_2015,lee_2015,yang_2016,he_2017,haneda_2018} provide useful insights but do not offer a means to evaluate proposed \mmwave full-duplex solutions since they provide neither a \mimo \si channel model nor adequate beam-based measurements; most of these were taken using horn or lens antennas, not phased ararys.
To evaluate beamforming-based \mmwave full-duplex solutions thus far, researchers have primarily used highly idealized channel models.
To address these shortcomings, a measurement campaign of \si at 28 GHz was conducted in \cite{roberts_att_angular,roberts_2021_att_measure} using a multi-panel 16$\times$16-element phased array platform.
In this campaign \cite{roberts_att_angular,roberts_2021_att_measure}, a spatial inspection of \si was conducted in an anechoic chamber by electronically sweeping the beams of the transmit phased array and receive phased array across a number of combinations in azimuth and elevation.
For each transmit direction and receive direction, \si power was measured, for a total of nearly 6.5 million measurements.
This work showed that \si indeed tends to be well above the noise floor---even with highly directional \mmwave beams---but select transmit-receive beam pairs coupled levels of \si below the noise floor without any additional cancellation. 
These measurements revealed large-scale trends based on steering direction, along with noteworthy small-scale phenomena when beams undergo small shifts (on the order of one degree).
The authors provide a statistical characterization of their measurements, allowing researchers to draw realistic realizations of \si and conduct statistical analyses.
A key takeaway from this work showed that a commonly-used idealized near-field channel model (i.e., the spherical-wave channel model \cite{spherical_2005}) is not a suitable one for practical \mmwave full-duplex systems. 
This motivates the need for a new measurement-backed channel model for \si in full-duplex \mmwave systems.


\begin{figure}[t]
    \centering
    \includegraphics[width=0.45\linewidth,height=\textheight,keepaspectratio]{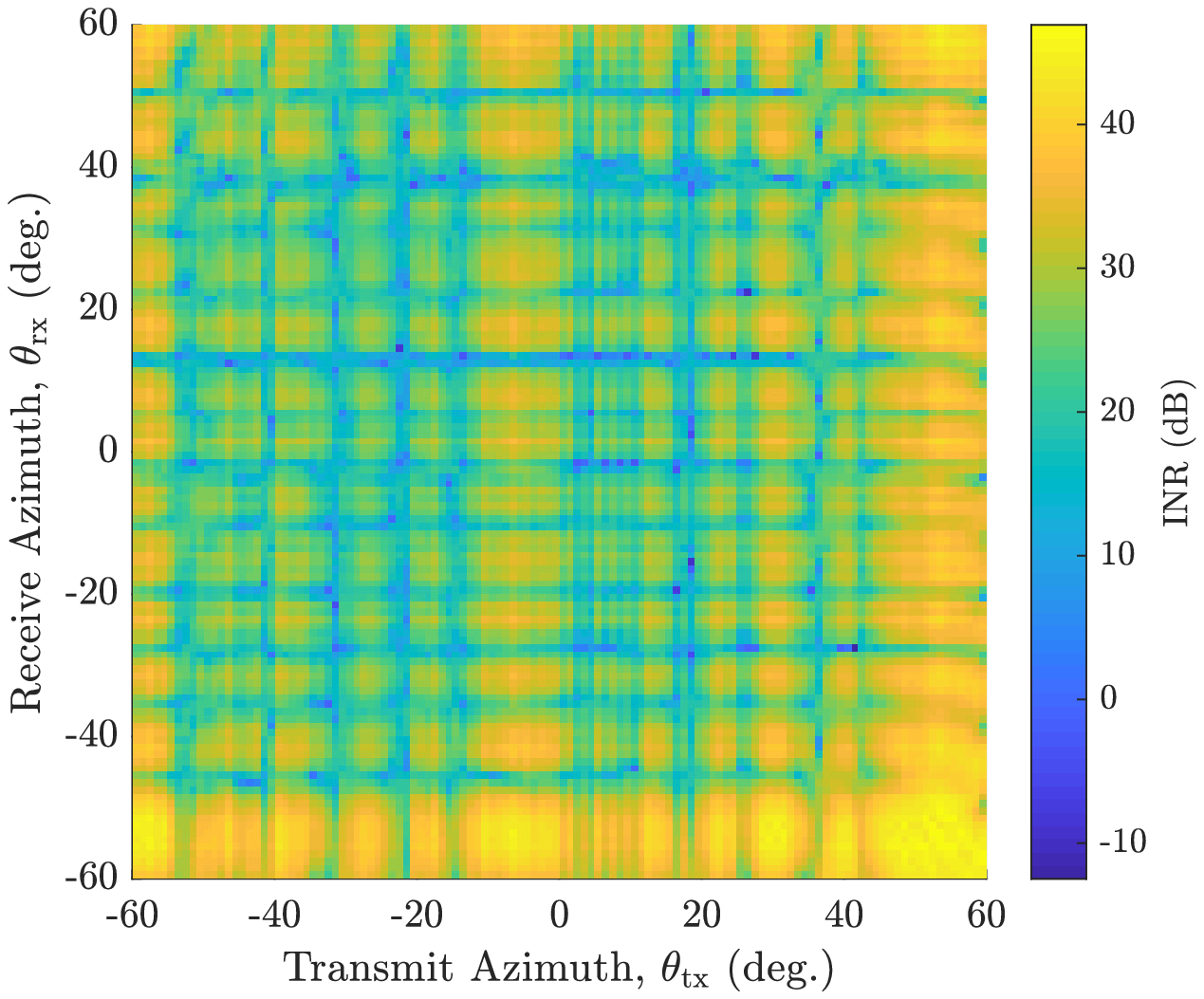}
    \qquad
    \includegraphics[width=0.45\linewidth,height=\textheight,keepaspectratio]{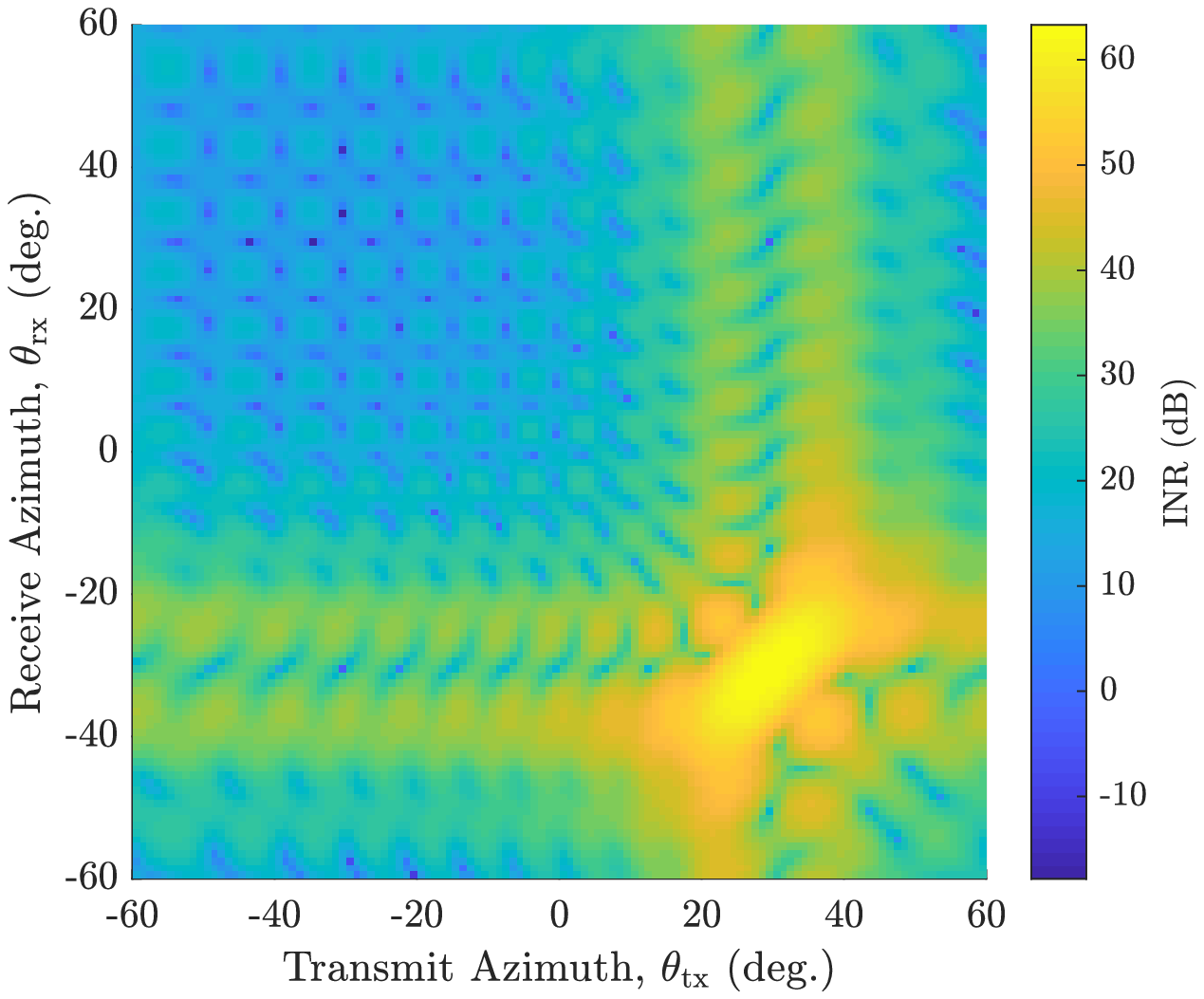}
    \caption{(left) The azimuth cut of \si measurements from \cite{roberts_att_angular}. (right) The simulated counterpart of (left) using a popular idealized near-field channel model for \si \cite{spherical_2005}. The stark difference between the two motivates the need for a new, measurement-backed \si channel model. Reproduced from \cite{roberts_att_angular} with permission.}
    \label{fig:model}
\end{figure}

\runinhead{Beam Selection for Full-Duplex mmWave Communication Systems.}
A particularly exciting observation from the measurements in \cite{roberts_att_angular} was that slightly shifting the transmit and/or receive beams at the full-duplex transceiver (on the order of one degree) could significantly reduce \si, often by $20$ dB or more.
This motivated the work of \cite{roberts_steer}, in which the authors propose the first beam selection methodology for full-duplex \mmwave communication systems.
Traditional beam selection typically selects beams that maximize \gsnr via codebook-based beam alignment measurements, as highlighted in \eqref{eq:beam-alignment}.
In \cite{roberts_steer}, the authors propose a measurement-driven beam selection methodology, called \steer, atop conventional beam alignment that incorporates \si into transmit and receive beam selection at a full-duplex \mmwave transceiver.


Suppose a full-duplex \bs serves a downlink user and uplink user as illustrated in \figref{fig:mmwave}.
Taking the perspective of a full-duplex \bs, let $\txdirsetcb$ be a set (a codebook) of $\Ntx$ candidate transmit beam steering directions (azimuth-elevation pairs) used during beam alignment, and let $\rxdirsetcb$ be a codebook of $\Nrx$ candidate receive beam steering directions defined analogously (e.g., see \figref{fig:alignment}).
\begin{align}
\txdirsetcb &= \braces{\parens{\thetatx\idx{i},\phitx\idx{i}} : i = 1,\dots,\Ntx} \\ 
\rxdirsetcb &= \braces{\parens{\thetarx\idx{j},\phirx\idx{j}} : j = 1,\dots,\Nrx}
\end{align}
Solving the following beam selection problems through conventional beam alignment yields initial beam selections at the full-duplex \bs.
\begin{align}
\thphtxiopt &= \argmax_{\thph \in \txdirsetcb} \ \snrtx\thph \\
\thphrxjopt &= \argmax_{\thph \in \rxdirsetcb} \ \snrrx\thph
\end{align}
While these initial beam selections may yield high \gpsnr, they are likely to couple high levels of \si, shown by measurements in \cite{roberts_att_angular}.
To identify transmit and receive beams that the \bs can use to deliver high \gpsnr and low \si, the authors of \cite{roberts_steer} propose \steer.

\begin{figure}[t]
    \centering
    \includegraphics[width=\linewidth,height=0.15\textheight,keepaspectratio]{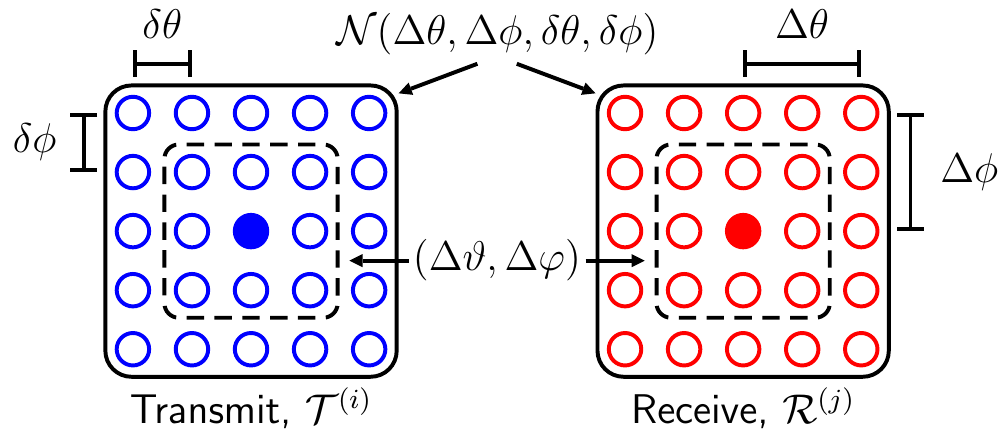}
    \caption{The spatial neighborhoods surrounding a given transmit direction and receive direction (shown as filled circles). The size of the neighborhoods is dictated by \nbr and their resolution by \nbrd. The sub-neighborhood \nbrv is relevant in problem \eqref{eq:problem-ii} \cite{roberts_steer}. Reproduced from \cite{roberts_steer} with permission.}
    \label{fig:neighborhood}
\end{figure}

\steer leverages the small-scale variability observed in the measurements of \cite{roberts_att_angular} to preserve high $\snrtx$ and high $\snrrx$ while reducing $\inrrx$.
To identify attractive steering directions for full-duplex operation, \steer measures the \si incurred when transmitting and receiving around the \textit{spatial neighborhoods} surrounding the initial transmit and receive steering directions, as described by \figref{fig:neighborhood}.
Quantifying the size of these spatial neighborhoods, let $\Deltatheta$ and $\Deltaphi$ be maximum absolute azimuthal and elevational deviations from the given transmit direction and receive direction. 
Discretizing these neighborhoods, let $\deltatheta$ and $\deltaphi$ be the measurement resolution in azimuth and elevation, respectively, which should not be larger than $\nbr$.
For instance, $\nbrd = \nbroneone$ while $\nbr = \nbrtwotwo$.
The spatial neighborhood $\setnbr$ surrounding a transmit/receive direction can be expressed using
the azimuthal neighborhood $\setnbrth$ and elevational neighborhood $\setnbrph$ defined as
\begin{align}
\setnbrth\nbrnbrdth &= \braces{m \cdot \deltatheta : m \in \brackets{-\floor{\frac{\Deltatheta}{\deltatheta}},\floor{\frac{\Deltatheta}{\deltatheta}}}} \\
\setnbrph\nbrnbrdph &= \braces{n \cdot \deltaphi : n \in \brackets{-\floor{\frac{\Deltaphi}{\deltaphi}},\floor{\frac{\Deltaphi}{\deltaphi}}}}
\end{align}
where $\floor{\cdot}$ is the floor operation and $\brackets{a,b} = \braces{a,a+1,\dots,b-1,b}$.
The complete neighborhood is the product of the azimuthal and elevational neighborhoods as
\begin{align}
\setnbr\nbrnbrd 
&= \braces{\thph : \theta \in \setnbrth\nbrnbrdth, \phi \in \setnbrph\nbrnbrdph}.
\end{align}
The spatial neighborhoods $\txdirsetmeas\idx{i\opt}$ and $\rxdirsetmeas\idx{j\opt}$ surrounding the transmit and receive directions output by conventional beam selection are respectively written as
\begin{align}
\txdirsetmeas\idx{i\opt} &= \thphtxiopt + \setnbr\nbrnbrd \\
\rxdirsetmeas\idx{j\opt} &= \underbrace{\thphrxjopt}_{\mathsf{initial~selection}} + \underbrace{\setnbr\nbrnbrd}_{\mathsf{neighborhood}}.
\end{align}

\begin{figure}[t]
    \centering
    \includegraphics[width=0.48\linewidth,height=0.235\textheight,keepaspectratio]{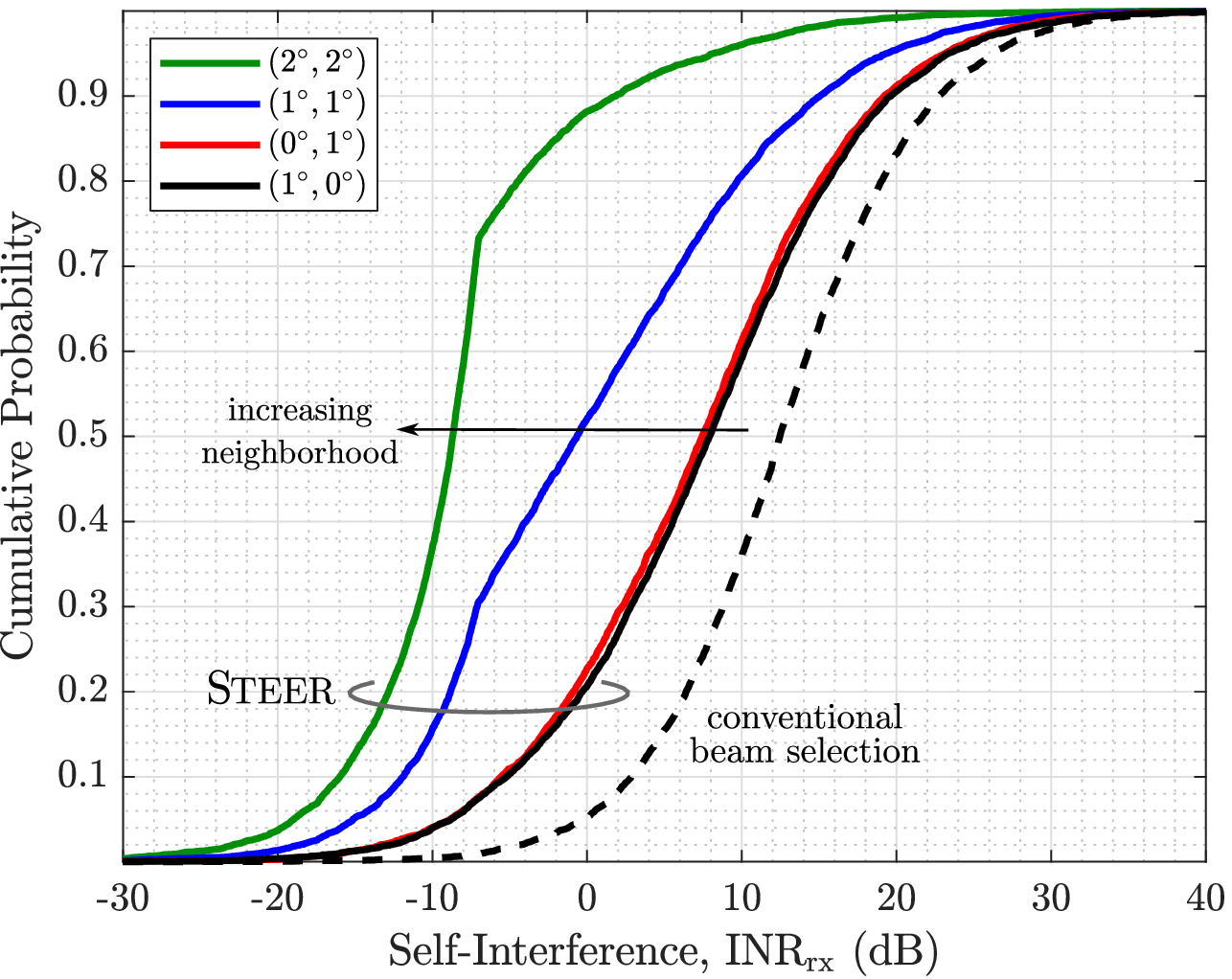}
    \quad
    \includegraphics[width=0.48\linewidth,height=0.2375\textheight,keepaspectratio]{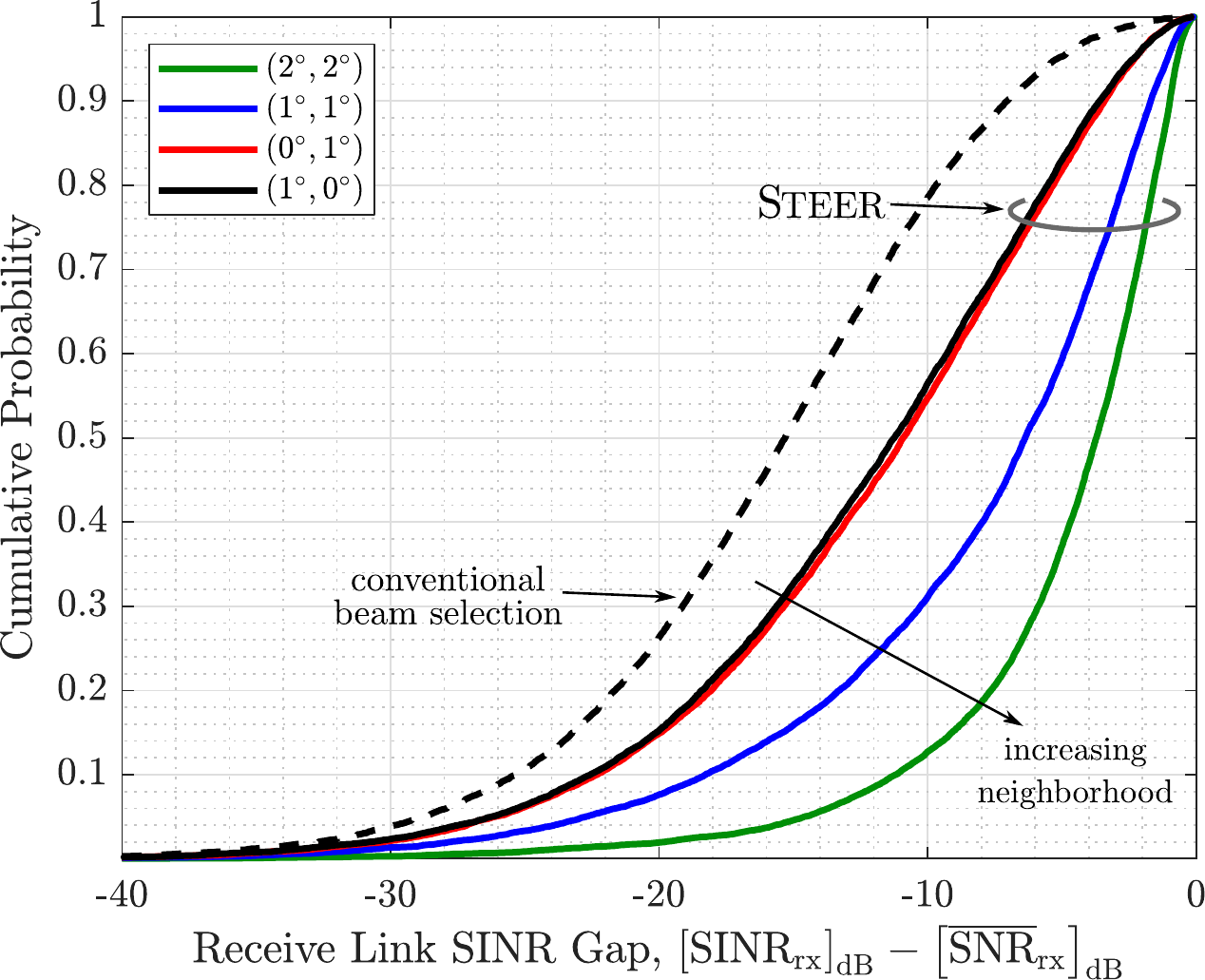}
    \caption{(left) The \gcdf of \inrrx for various neighborhood sizes \nbr. (right) The \gcdf of the gap between $\sinrrx$ and its upper bound $\snrrxbar$ for various neighborhood sizes $\nbr$. \steer reliably reduces $\inrrx$, as evident in (left), while maintaining high beamforming gain, shifting $\sinrrx$ closer to $\snrrxbar$ as shown in (right) \cite{roberts_steer}. Reproduced from \cite{roberts_steer} with permission.}
    \label{fig:cdf-steer}
\end{figure}

Let $\inrrx\thphtxrx$ be the receive link \ginr due to \si when transmitting toward $\thphtx$ and receiving toward $\thphrx$ at the full-duplex \bs.
\steer solves the following beam selection problem to net a transmit direction $\thphtxopt$ and receive direction $\thphrxopt$ that the full-duplex transceiver will use.%
\begin{subequations}\label{eq:problem-ii}%
\begin{align} 
\thphtxopt, \thphrxopt = 
\argmin_{\substack{\thphtx\\\thphrx}} & \min_{\parens{\Deltavartheta,\Deltavarphi}} \ \Deltavartheta^2 + \Deltavarphi^2 \\
\st \ 
& \inrrx\thphtxrx \leq \maxop{\inrrxthresh,\inrrxmin} \label{eq:target-inr-constraint} \\
& \thphtx \in \thphtxiopt + \setnbr\nbrvnbrd \\
& \thphrx \in \thphrxjopt + \setnbr\nbrvnbrd \\
& 0 \leq \Deltavartheta \leq \Deltatheta, \ 0 \leq \Deltavarphi \leq \Delta\phi
\end{align}
\end{subequations}
Here, $\inrrxmin$ is the minimum \ginr over the entire $\nbr$ spatial neighborhood and $\inrrxthresh$ is an \ginr target the system aims for.
By constraining the distance $\Delta\vartheta^2 + \Delta\varphi^2$, \steer minimizes the deviation of the selected beams make from the initial selections and thus can preserve high $\snrtx$ and $\snrrx$.
Reducing $\inrrx$, therefore, leads to \gsinr improvements over the initial beam selections.
In \cite{roberts_steer}, the authors present an algorithm to solve problem \eqref{eq:problem-ii} with a minimal number of \si measurements.
Evaluation of \steer with 28 GHz phased arrays highlights its ability to reduce \si while preserving high \gpsnr, courtesy of noteworthy variability of \si over small spatial neighborhoods.
This can be seen in \figref{fig:cdf-steer}, which shows \steer's potential as a full-duplex solution without any supplemental analog or digital \sic.
\si can be reliably reduced with \steer to below the noise floor with $\nbr = \nbrtwotwo$, and by preserving \gsnr while doing so, it can increase \gsinr toward its upper bound.

\section{A Look Ahead: What is the Future of Full-Duplex?} \label{sec:look-ahead}

We conclude this chapter by highlighting several key topics that need further research and development from engineers in industry and academia to advance and mature full-duplex technology.

\runinhead{Real-time, deployment-ready full-duplex solutions.}
Most full-duplex solutions are validated in simulation, lab settings, or controlled environments.
Moreover, most evaluations ignore the overhead associated with configuring a full-duplex solution, which is a key hurdle in practical deployments.
In practice, the radio resources consumed to configure a full-duplex solution must not outweigh the gains it offers.
It is essential that full-duplex solutions are designed and evaluated with real-time deployments in mind---with strict overhead requirements, with the ability to adapt to dynamic environments, and with minimal form-factors and power consumption.


\runinhead{Further advancement of full-duplex \mmwave and \acrlong{thz} systems.}
While increased attention has been devoted to the research and development of full-duplex \mmwave systems, there remain plenty of open problems that need addressing before such systems are brought to life.
Continuing to develop means to mitigate \si is certainly welcome, along with characterizing the gains full-duplex can offer \mmwave networks and prototyping proofs-of-concept.
In addition, identifying and creating applications that are particularly beneficial from full-duplex would offer new directions and requirements for its solutions.
For instance, the role of full-duplex in \gls{ris} applications has been investigated recently as it poses a means to better use \gls{ris} resources and aid in the cancellation of \si \cite{liu_ris_2021,himal_ris_icc_2022}.
Finally, exploring full-duplex \acrlong{thz} systems, their applications, and how solutions for such may differ from those at \mmwave would be valuable future work.

\runinhead{Further advancement of machine learning to enable full-duplex.}
To supplement existing work, the prospects of using machine learning for full-duplex are still quite open-ended, especially beyond digital \sic. 
Using machine learning to configure and adaptively update analog \sic filters, for instance, or to configure beamformers that mitigate \si in full-duplex \mmwave systems are topics that have yet to be fully explored.
In addition, machine learning may be able to reduce the effects of transceiver impairments in full-duplex through digital predistortion.
There are also the prospects of using machine learning to intelligently schedule users and proactively manage cross-link interference within full-duplex networks. 

\runinhead{Network-level studies comparing full-duplex to other duplexing technologies.}
To justify the deployment of wireless networks equipped with full-duplex, it is paramount that researchers conduct studies that prove its network-level gains over traditional multiplexing strategies, such as \tdd, \fdd, and \acrlong{sdma}, as well as the recently-proposed \gls{xdd} \cite{xdd}.
This has been examined fairly extensively for sub-6 GHz wireless networks but less so for \mmwave networks and applications of \iab, in particular. 

\runinhead{Full-duplex in joint communication and sensing systems.}
Joint communication and sensing is expected to play an important role in the next evolution of wireless networks.
Full-duplex solutions have the opportunity to facilitate high-fidelity sensing while transmitting by eliminating  \si \cite{jcas_wcm_2021,xiao_jsac_2022}. 
Sensing information may be used to directly improve communication performance \cite{ali_leveraging_2020} or for higher-level applications.
In addition, there are opportunities for full-duplex to enable the sensing and jamming of eavesdroppers to establish more secure communication \cite{wang_coml_2022}.
Finally, the relationship between \si channel estimation and environmental sensing poses an opportunity for the two to supplement and/or justify one another.
For instance, accurate sensing of the environment may yield an estimate of the \si channel and, in turn, enable full-duplex operation.



\runinhead{Integrating full-duplex into wireless standards.}
Wireless networks of today have been built on decades of a half-duplex assumption at each transceiver.
Full-duplex offers immediate upgrades to a transceiver but, to effectively make use of this powerful capability, wireless networks must support it. 
This motivates the need for research on seamlessly adopting full-duplex operation into existing wireless standards.
Work items and studies in the \gls{3gpp} continue to investigate the merits and standardization of full-duplex in cellular systems, especially in the context of \iab in Releases 17 and 18 \cite{3GPP_IAB_2}.

\begin{acknowledgement}
We would like to thank Jeffrey G.~Andrews, Sriram Vishwanath, Aditya Chopra, Thomas Novlan, Manan Gupta, Rajesh K.~Mishra, and Hardik B.~Jain for the discussions, feedback, and collaborations that contributed to the preparation of this chapter.
The work of Ian P. Roberts was supported by the U.S.~National Science Foundation under Grant DGE-1610403.
\end{acknowledgement}

\bibliographystyle{bibtex/IEEEtran}
\bibliography{bibtex/IEEEabrv,refs}

\end{document}